\documentclass[twocolumn,10pt,cleanfoot]{ihmtc}
\usepackage[T1]{fontenc}
\usepackage{balance}
\usepackage{graphicx}
\usepackage{grffile}
\usepackage{epstopdf}
\usepackage{mathtools}
\usepackage{subcaption}
\usepackage{cuted}
\usepackage{etoolbox}
\usepackage{blindtext}
\usepackage{float}

\AfterEndEnvironment{strip}{\leavevmode}

\graphicspath{{Figures/}}

\conffullname{the 25th National and 3rd International ISHMT-ASTFE\\
		Heat and Mass Transfer Conference (IHMTC-2019),}

\confdate{December 28-31}
\confyear{2019}
\confcity{IIT Roorkee, Roorkee}
\confcountry{India}
\usepackage{gensymb}
\papernum{IHMTC2019-CFD-808}
\usepackage{lipsum}

\title{Instabilities of Decelerating Flow}

\author
{Ahire Swapnil Ashok$^{a,\ast}$, Manu K.V. $^{a}$\\
\\
\normalsize{$^{a}$Department of Aerospace Engineering, Indian Institute of Space Science and Technology, Thiruvananthapuram, India}\\
\\
\normalsize{$^\ast$Corresponding author's email:  swapnilahire15@gmail.com}
}	

\begin{document}

\maketitle 

\begin{abstract}
{
This paper numerically investigates the instability characteristics of decelerating flows. The flow dynamics and temporal evolution of coherent structures in a diverging section with mild spatial pressure gradient are analyzed using numerical experiments with \textit{Incompact3d} code. The unsteady nature of flow and adverse pressure gradient is the reason for inflectional velocity profiles, results into boundary layer separation and instability with reverse flow and later on it leads to vortex formation. Vortex formation time is found by vortex Reynold's number. Time of first vortex formation, non-dimensional vortex formation times with Reynold's number etc. are validated with an experimental results.
}
\end{abstract}
\begin{nomenclature}
	\entry{$U_e$}
		{ Velocity at the edge of the boundary layer.}
	\entry{$\mathbf{u} (\mathbf{x},t)$} 
		{Velocity field}
	\entry{$p (\mathbf{x},t)$} 
		{Pressure field}
	\entry{$\mathbf{f} (\mathbf{x},t)$} 
		{Forcing field to impose immersed boundary method}	
	\entry{$U$} {Inlet velocity}
	\entry{$U_0$} {Maximum inlet velocity}		
	\entry{$u_p$} {Piston Velocity}
\entry{$h$} {Half height of channel}
\entry {$y$}{Distance from center of channel towards the wall}
\entry {$c$}{$=\frac{y}{h} $}
\entry {$v_{n h}$} {Roots of $tan (v) = v\text{ and } v_{nh}=1,2, \dots \dots \dots \infty$}
\entry {d} {the deceleration rate at the inlet}
\entry {$t_s$} {time at which separating occurred}
\entry {$t_v$} {time at which first vortex is formed}
\entry {$\delta_s$} {Boundary layer thickness at time $t_s$}
\entry {$\Delta U_s$} {($U_{max}-U_{min}$) at time $t_s$}
\entry {$Re_{\delta s}$} {Reynolds number at time $t_s$}

\end{nomenclature}

\section*{INTRODUCTION}
In unsteady flow, boundary layer separates and leads to instability of flow if necessary conditions are present for separation and instability. Unsteady flow and mild pressure gradient is cause flow to separate and to form instability which will probably lead to vortex formation. Unsteady flow is naturally occurred like flow over wings of bird, blood flow in heart and in case of applications like turbine blades,helicopter blades, pumps etc.

\par	
Momentum equation for 2 dimensional unsteady boundary layer in flow direction is given by,
\begin{equation}
\frac{\partial u}{\partial t}+u \frac{\partial u}{\partial x}+v \frac{\partial u}{\partial y}=-\frac{1}{\rho} \frac{\partial P}{\partial x}+v \frac{\partial^{2} u}{\partial y^{2}}
\end{equation}
where, \textit{x} and \textit{y} are boundary layer coordinates
 t is time, \textit{u} and \textit{v} are components of velocity in x and y direction

Pressure gradient term can be written with two components which is given as,

\begin{equation}
-\frac{1}{\rho} \frac{\partial P}{\partial x}=\frac{\partial U_{e}}{\partial t}+U_{e} \frac{\partial U_{e}}{\partial x}
\label{eq:decomp}
\end{equation}

\par
On right hand side of equation (\ref{eq:decomp}), $\frac{\partial U_{e}}{\partial t}$ is called temporal component and $U_{e} \frac{\partial U_{e}}{\partial x}$ is called spatial component. Temporal term gives the acceleration or deceleration of the free-stream and spatial component gives the convective acceleration of the free stream. Most of the studies had been done in straight tubes or channels where spatial component is absent but in case of biological systems like pulsating flow in arteries  and application based engineering system, the spatial term is present. Blood flow in arteries \cite{ku1997blood} and dynamic stall \cite{buchner2015measurements} are the examples where pulsating flow is present with spatial term.
\par
Analytical solutions are obtained for laminar unidirectional flow. If pressure gradient is known then solution for sinusoidal variation of pressure gradient is given by Uchida \cite{uchida1956pulsating}. D. Das and Arakeri \cite{das2000unsteady} has proposed a procedure for unsteady laminar analytical solution for infinitely long pipe with circular and  infinitely long 2-D channel provided that flow rate is given as function of time, measured experimentally or by piston controlled motion.

\par 
Boundary layer separation can be seen as moving away of boundary layer away from the surface and may results in breaking of the boundary layer assumption. If adverse pressure gradient is present, inflectional profile can be develop. It might lead to the instability at low Reynolds number. Instability results into the vortex formation\cite{das2016instabilities}. Gaster and Michael \cite{gaster1966structure} have performed experiment to understand laminar steady bubble with adverse pressure gradient on the flat plate. Later on, experiments performed by Gaster and Michael are validated numerically by Ripley et al. \cite{ripley1993unsteady} and provided better understanding of separation bubble.

\section*{NUMERICAL METHODS}
\subsection*{Governing Equations}
Governing equations is forced incompressible Navier-Stokes equation in skew-symmetric form is given below,
		\vspace{-1mm}												
\begin{equation}
\frac{\partial \mathbf{u}}{\partial t}=-\nabla p-\frac{1}{2}[\mathbf{\nabla}(\mathbf{u} \otimes \mathbf{u})+(\mathbf{u} . \nabla) \mathbf{u}]+v \nabla^{2} \mathbf{u}+\mathbf{f}
\label{eq:ge1}
\end{equation}
		\vspace{-10mm}												
\begin{equation}
\nabla . \mathbf{u}=0
\label{eq:ge2}
\end{equation}
Governing equations are solved with \textit{Incompac3d} code \cite{laizet2009high}, in which domain is spatially discretized with sixth order accurate compact finite difference schemes and time marching is done with second order accurate Adams-bashforth scheme.
Immersed boundary method is imposing the forcing field in the equation (\ref{eq:ge1}).

\subsection*{Computational Domain}
Numerical experiments are done on diffuser with an angle of $7.4^{\circ}$ as shown in figure (\ref{domain}). As diffuser angle is less, it has mild spatial gradient. There should not be sudden change in pressure gradient, hence small curvature is provided to join inclined and straight sections. The domain is taken similar as S.P Das \cite{das2016instabilities} in order to validate.

{\begin{figure}[h]
		\centering
		\includegraphics[width=1\linewidth]{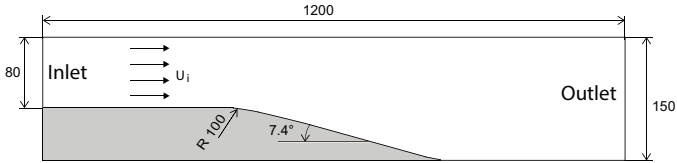}
		\vspace{-20mm}				
		\caption{ COMPUTATIONAL DOMAIN FOR SIMULATION (ALL DIMENSIONS ARE IN MM)}
		\label{domain}
		\vspace{-10mm}						
	\end{figure}
\begin{table*}[h]
	\centering
	\begin{tabular}{|c|c|c|c|c|c|c|c|c|c|c|c|c|c|c|}
		\hline
		& \multicolumn{5}{c|}{Input Conditions} & \multicolumn{5}{c|}{Simulation} & \multicolumn{4}{c|}{Experiment} \\ \hline
		Case & \begin{tabular}[c]{@{}c@{}}$U_0$\\ ($cm/s$)\end{tabular} & \begin{tabular}[c]{@{}c@{}}$d$\\ ($cm/s^2$)\end{tabular} & \begin{tabular}[c]{@{}c@{}}$t_0$\\ ($s$)\end{tabular} & \begin{tabular}[c]{@{}c@{}}$t_1$\\ ($s$)\end{tabular} & \begin{tabular}[c]{@{}c@{}}$t_2$\\ ($s$)\end{tabular} & \begin{tabular}[c]{@{}c@{}}$t_s$\\ ($s$)\end{tabular} & \begin{tabular}[c]{@{}c@{}}$t_v$\\ ($s$)\end{tabular} & \begin{tabular}[c]{@{}c@{}}$\delta_s$\\ ($mm$)\end{tabular} & \begin{tabular}[c]{@{}c@{}}$\Delta U_s$\\ ($m/s$)\end{tabular} & $Re_{\delta_s}$ 
		& \begin{tabular}[c]{@{}c@{}}$t_s$\\ ($s$)\end{tabular} & \begin{tabular}[c]{@{}c@{}}$t_v$\\ ($s$)\end{tabular} & \begin{tabular}[c]{@{}c@{}}$\delta_s$\\ ($mm$)\end{tabular}  & $Re_{\delta_s}$\\ \hline
		1 & 13.72 & 2.29 & 0.6 & 2.0 & 8.0 & 2.50 & 4.80 & 6.38 & 0.1263 & 805 & 2.50 & 5.2 & 5.70 & 733 \\ \hline
		2 & 13.72 & 4.57 & 0.6 & 3.5 & 6.5 & 3.50 & 5.50 & 8.01 & 0.1367 & 1095 & 3.50 & 5.2 & 7.18  & 1000 \\ \hline
		3 & 18.30 & 4.57 & 0.8 & 1.0 & 5.0 & 1.70 & 3.50 & 4.95 & 0.1489 & 737 & 1.75 & 3.40 & 4.78 & 719\\ \hline
		4 & 13.72 & 13.72 & 0.6 & 2.0 & 3.0 & 2.05 & 4.40 & 5.41 & 0.1317 & 713 & 2.05 & 4.27 & 5.07 & 672 \\ \hline
		5 & 27.45 & 32.03 & 1.2 & 2.0 & 2.83 & 2.00 & 2.50 & 5.41 & 0.2651 & 1434 & 2.05 & 2.50 & 5.07 & 1355 \\ \hline
	\end{tabular}
	\caption{INPUT CONDITIONS WITH COMPARISON OF PARAMETERS BETWEEN SIMULATION AND EXPERIMENTS BY S.P DAS \cite{das2016instabilities}}
	\label{tab:sim}
	\vspace{-10mm}
\end{table*}
\subsection*{Boundary Conditions}
Trapezoidal velocity variation with time is provided at the inlet of domain. Variation of velocity with time is as shown in Figure(\ref{trap}). 

{\begin{figure}[htb!]
		\centering
		\includegraphics[width=0.61\linewidth]{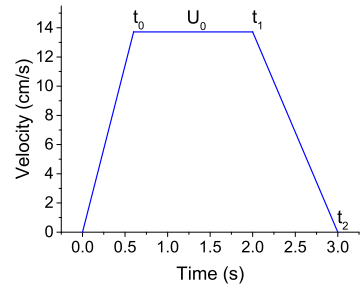}
		\caption{ VARIATION OF INLET VELOCITY $U$ WITH TIME AND $U_0$ IS MAXIMUM VELOCITY}
		\label{trap}
		\vspace{-7mm}												
	\end{figure}
	Analytical solution provided by D. Das \cite{das2000unsteady} for the trapezoidal variation of the mean velocity with time is defined using the following expressions:
	\begin{equation}
	\begin{aligned} u_{p}(t) &=\frac{U_{0} t}{t_{o}} & \text { for } 0 \leq t \leq t_{o} \\ &=U_{0} & \text { for } t_{o} \leq t \leq t_{1} \\ &=\frac{U_{0}\left(t_{2}-t\right)}{\left(t_{2}-t_{1}\right)} & \text { for } t_{1} \leq t \leq t_{2} \\ &=0  & \text { for } t_{2} \leq t \leq \infty \end{aligned}
	\label{eq:trap}
	\end{equation}
		Solution for trapezoidal piston motion for channel flow at different times can be given by,

\begin{itemize}
	\item $0 \leq t \leq t_{o}$
	\begin{equation}
	\begin{aligned} \frac{U}{U_{p}}=& \frac{1}{t_{o}}\left(\frac{3 t}{2}\left(1-c_{h}^{2}\right)-\frac{h^{2}}{40 \nu}\left(5 c_{h}^{4}-6 c_{h}^{2}+1\right)\right) \\ &-\frac{2 h^{2}}{\nu t_{o}} \sum_{n h=1}^{\infty} e^{-v_{n h}^{2} \nu t / h^{2}}\left[\frac{\cos \left(c_{h} v_{n h}\right)-\cos \left(v_{n h}\right)}{v_{n h}^{3} \sin \left(v_{n h}\right)}\right] \end{aligned}
	\label{eq:trap1}
	\end{equation}
	\item $t_{o} \leq t \leq t_{1}$
	\begin{equation}
	\begin{aligned} \frac{U}{U_{p}}=& \frac{3}{2}\left(1-c_{h}^{2}\right)-\frac{2 h^{2}}{\nu t_{O}} \sum_{n h=1}^{\infty}\left(e^{-v_{n h}^{2} \nu / h^{2} ) t}\left(-e^{-v_{n h}^{2} \nu / h^{2}}\right)\left(t-t_{o}\right)\right) \\ & \times\left[\frac{\cos \left(c_{h} v_{n h}\right)-\cos \left(v_{n h}\right)}{v_{n h}^{3} \sin \left(v_{n h}\right)}\right] \end{aligned}
	\label{eq:trap2}
	\end{equation}
	\item $t_{1} \leq t \leq t_{2}$
	\begin{equation}
	\begin{array}{c}{\frac{U}{U_{p}}=\frac{3}{2}\left(1-c_{h}^{2}\right)\left(\frac{t_{2}-t}{t_{2}-t_{1}}\right)+\frac{1}{t_{2}-t_{1}} \frac{h^{2}}{40 \nu}\left(5 c_{h}^{4}-6 c_{h}^{2}+1\right)} \\ {-\frac{2 h^{2}}{\nu} \Sigma_{n h=1}^{\infty}\left(\frac{e^{\left(-v_{n h}^{2} \nu / h^{2}\right) t}-e^{\left(-v_{n h}^{2} \nu / h^{2}\right)\left(t-t_{o}\right)}}{t_{o}}\right.} \\ {-\frac{e^{\left(-v_{n h}^{2} \nu / h^{2}\right)\left(t-t_{1}\right)}}{t_{2}-t_{1}} )\left[\frac{\cos \left(c_{h} v_{n h}\right)-\cos \left(v_{n h}\right)}{v_{n h}^{3} \sin \left(v_{n h}\right)}\right]}\end{array}
	\label{eq:trap3}
	\end{equation}
	\item $t_{2} \leq t \leq \infty$
	\begin{equation}
	\begin{aligned} \frac{U}{U_{p}}=&-\frac{2 h^{2}}{\nu} \sum_{n h=1}^{\infty}\left(\frac{e^{\left(-v_{n h}^{2} \nu / h^{2}\right) t}-e^{\left(-v_{n h}^{2} \nu / h^{2}\right)\left(t-t_{o}\right)}}{t_{o}}\right.\\ &+\frac{e^{\left(-v_{n h}^{2} \nu / h^{2}\right)\left(t-t_{2}\right)}-e^{\left(-v_{n h}^{2} \nu / h^{2}\right)\left(t-t_{1}\right)}}{x_{2}-t_{1}} ) \\ & \times\left[\frac{\cos \left(c_{h} v_{n h}\right)-\cos \left(v_{n h}\right)}{v_{n h}^{3} \sin \left(v_{n h}\right)}\right] \end{aligned}
	\label{eq:trap4}
	\end{equation}
At outlet of the domain, convective boundary condition is applied which is given by equation($\ref{eq:convective}$) and no-slip boundary condition is imposed on top as well as bottom wall.
Initially, a grid independence study has been performed using three grid configurations which are (a) 729 $\times$ 257, (b) 1025 $\times$ 433 and (c) 1459 $\times$ 513. It is observed that the differences in flow variables between grid (b) and (c) are negligible but grid (c) is used for the rest of the simulations with time step of 0.0001 seconds. Mesh is stretched at center of the domain. 
\end{itemize}

\begin{equation}
\vspace{-10mm}
u_{t}+u \cdot \nabla u = 0
\label{eq:convective}
\end{equation}

{\begin{figure}[htb!]
		\centering									
		\includegraphics[width=1\linewidth]{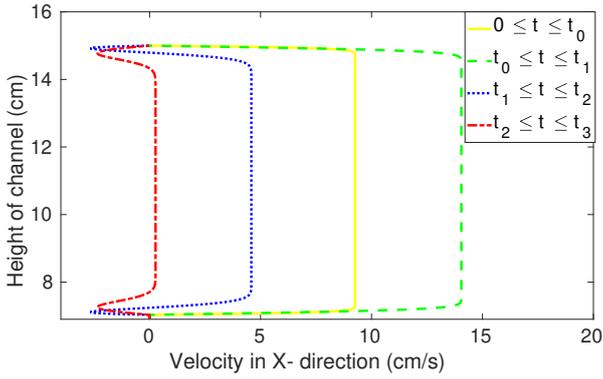}			
		\caption{VELOCITY PROFILES IMPOSED AT INLET BY ANALYTICAL SOLUTION PROVIDED BY D. DAS \cite{das2000unsteady}}
		\label{fig:ts2}
		\vspace{-5mm}											
	\end{figure}

\section*{RESULTS AND DISCUSSION}
For flow in diverging channel, instability arises due to inflectional velocity profile. Instability is initially observed when inlet velocity is under deceleration or after piston stops. Numerical experiments were conducted for five cases. Deceleration of inlet is varied from 2.29 $m/s^2$ to 32.03 $m/s^2$. Maximum velocity given at the inlet is 27.45 $m/s$ and minimum velocity is 13.72 $m/s$. Comparison between simulation results and experimental results is given in Table(\ref{tab:sim}).
\par
Time of separation ($t_s$) is considered as, occurrence of reverse velocity at the grid point adjacent to the wall. To find the separation, at the time step on which data is stored, velocity at the adjacent grid point to the wall is checked. Once the point at which first separation is occurring is found then velocity in x-direction will be plotted against time.
\par
Time of first vortex formation time is found by vortex reynolds number ($Re_v$). Consider a arbitrary closed loop contour ($C$) of area ($A$) and small element on contour $dl$. Circulation about contour $C$ is defined as,
\begin{equation}
\Gamma=\oint_{\mathcal{C}} \mathbf{u} \cdot dl=\int_{\mathcal{A}}(\nabla \times \mathbf{u}) \cdot \widehat{\mathbf{z}} d A=\int_{\mathcal{A}} \omega d A
\label{eq:tv2}
\end{equation}

A vortex Reynolds number ($Re_v$) \cite{doligalski1994vortex} can be defined in terms of circulation ($\Gamma$) and kinematic viscosity ($\nu$). It is given by,
\begin{equation}
Re_v=\frac{\Gamma}{2 \pi \nu}
\label{eq:rev1}
\end{equation}

By combining equation (\ref{eq:tv2}) and (\ref{eq:rev1}), vortex reynolds number can be written as,

\begin{equation}
Re_v=\frac{\int_{\mathcal{A}} \omega d A}{2 \pi \nu}
\label{eq:rev2}
\end{equation}

Above equation (\ref{eq:rev2}) is used for finding vortex Reynolds number for all the cases. Vortex Reynolds number are computed on diverging section of the channel starting from the time of separation. The peak point on the plot of vortex reynolds number against time is taken as time of first vortex formation ($t_v$).

{\begin{figure}[h]
		\centering
		\begin{subfigure}{0.22\textwidth}
			\centering
			\includegraphics[width=1\linewidth]{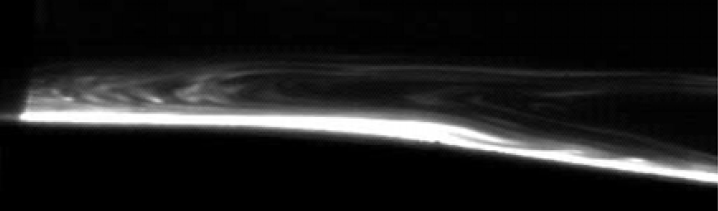}
		\end{subfigure}%
		\begin{subfigure}{0.25\textwidth}
			\centering
			\includegraphics[width=1\linewidth]{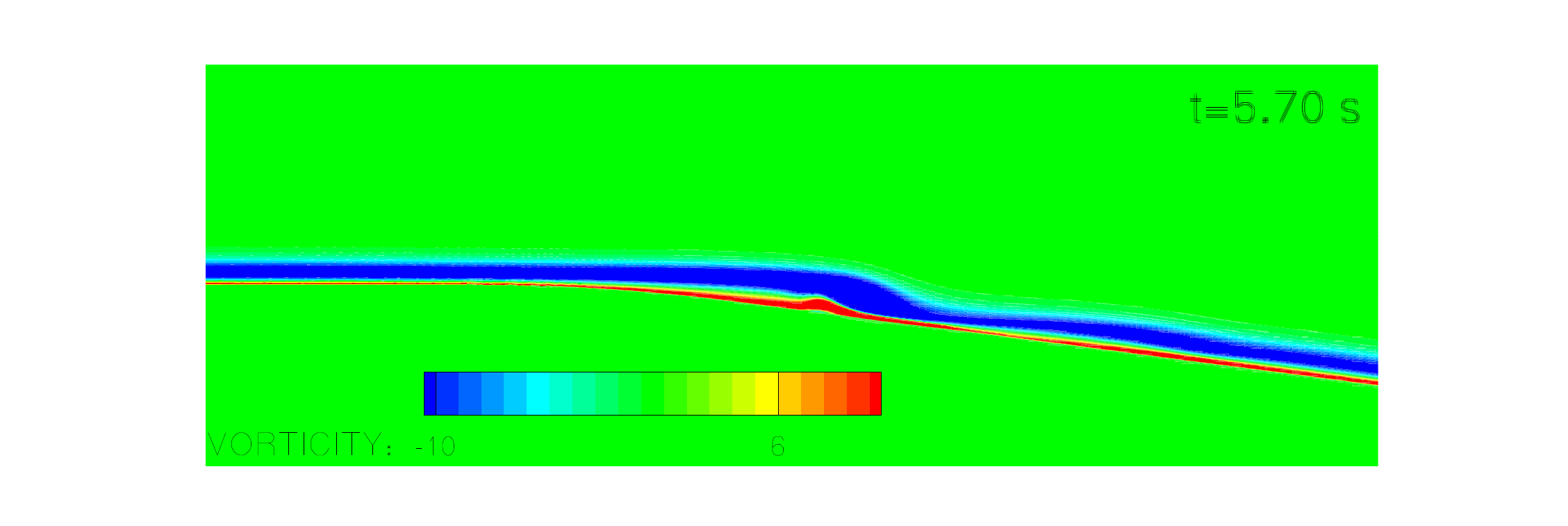}
		\end{subfigure}
		\begin{subfigure}{0.22\textwidth}
			\centering
			\includegraphics[width=1\linewidth]{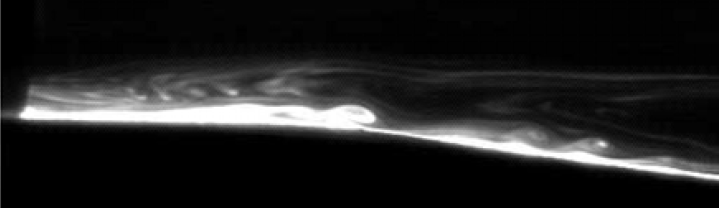}
		\end{subfigure}%
		\begin{subfigure}{0.25\textwidth}
			\centering
			\includegraphics[width=1\linewidth]{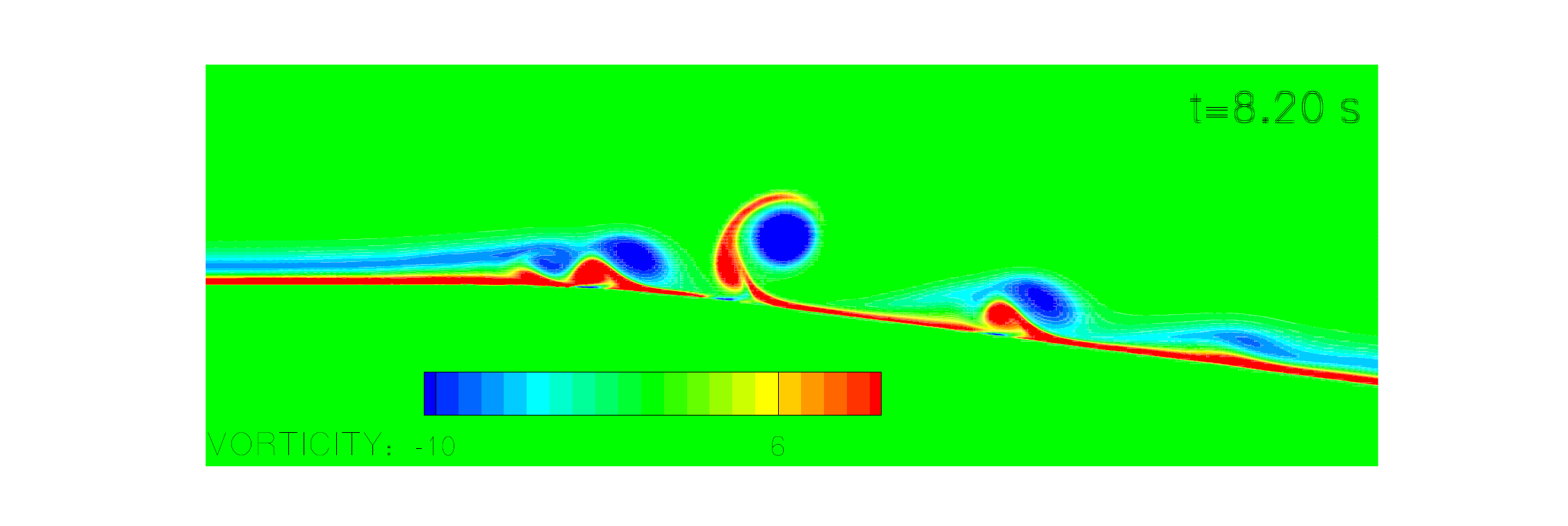}
		\end{subfigure}
		\begin{subfigure}{0.22\textwidth}
			\centering
			\includegraphics[width=1\linewidth]{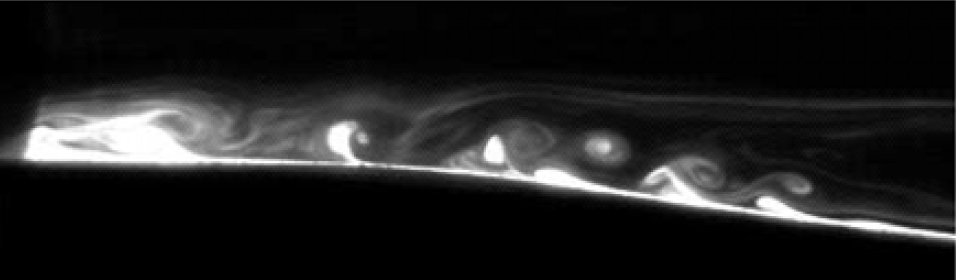}
		\end{subfigure}%
		\begin{subfigure}{0.25\textwidth}
			\centering
			\includegraphics[width=1\linewidth]{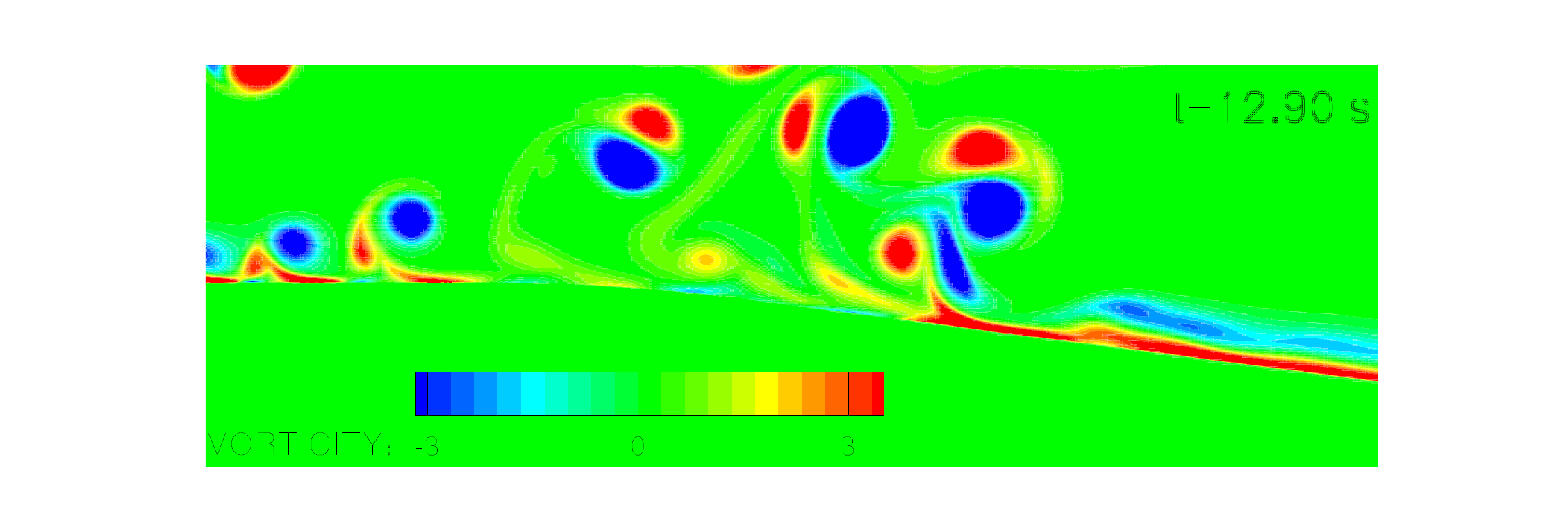}
		\end{subfigure}
		\begin{subfigure}{0.22\textwidth}
			\centering
			\includegraphics[width=1\linewidth]{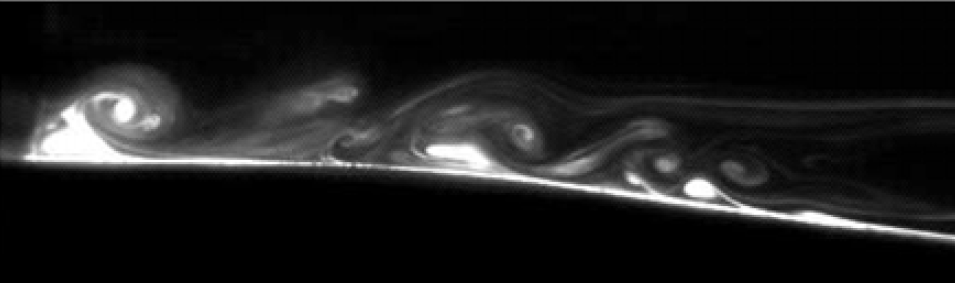}
		\end{subfigure}%
		\begin{subfigure}{0.25\textwidth}
			\centering
			\includegraphics[width=1\linewidth]{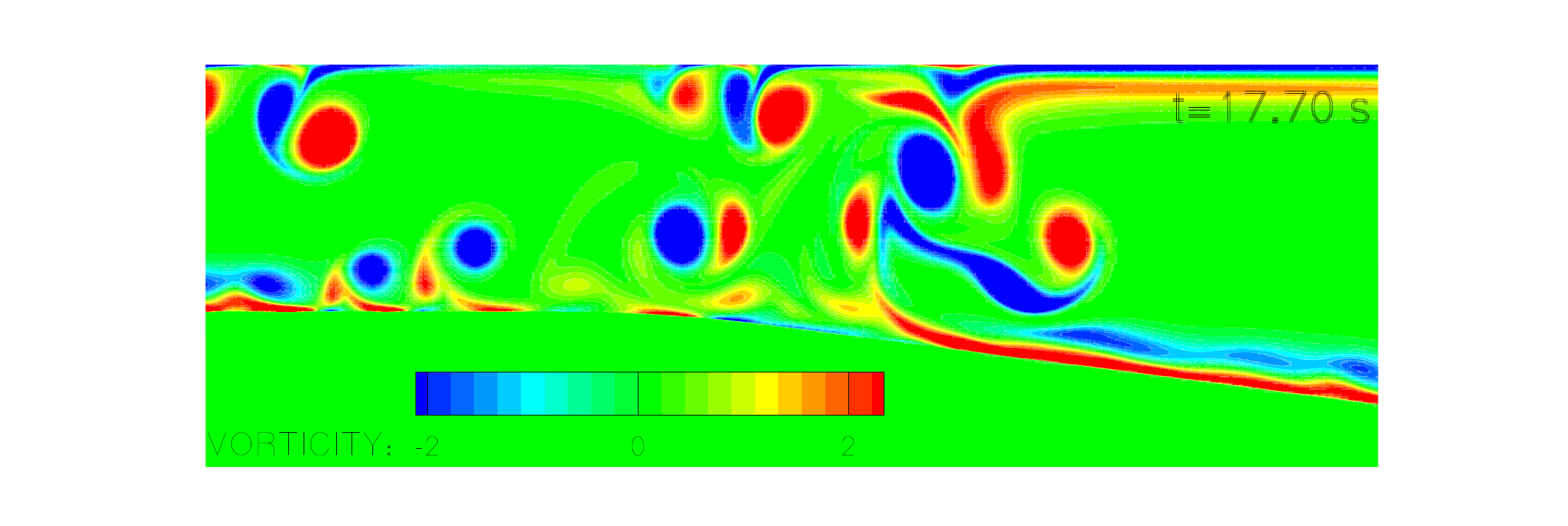}
		\end{subfigure}
		\caption[Case 1 Contour Plot] {CASE 1 $t_0=0.6 s, t_1=2.0 s, t_2=8.0 s,U_0=13.72 cm/s, Re_{\delta} \simeq 805 $ \\
	COMPARISON BETWEEN EXPERIMENTAL RESULT FROM DAS ET. AL \cite{das2016instabilities}  AND SIMULATION VORTICITY CONTOUR PLOT }
\label{fig:case4_1}
\end{figure}}

{\begin{figure}[h]
		\centering
		\begin{subfigure}{0.22\textwidth}
			\centering
			\includegraphics[width=1\linewidth]{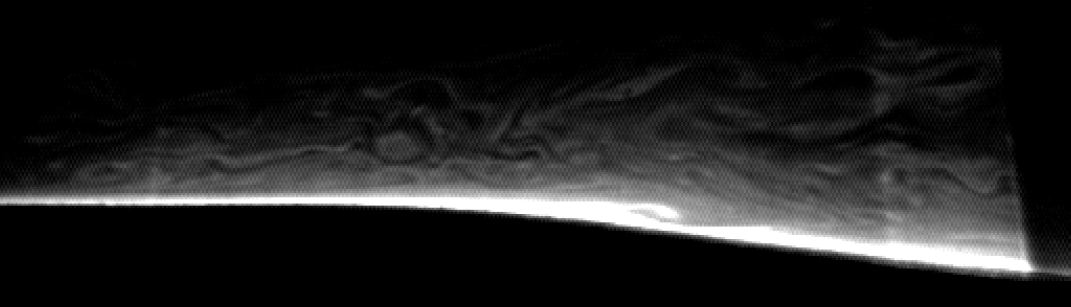}
		\end{subfigure}%
				\vspace{2mm}
		\begin{subfigure}{0.22\textwidth}
			\centering
			\includegraphics[width=0.9\linewidth]{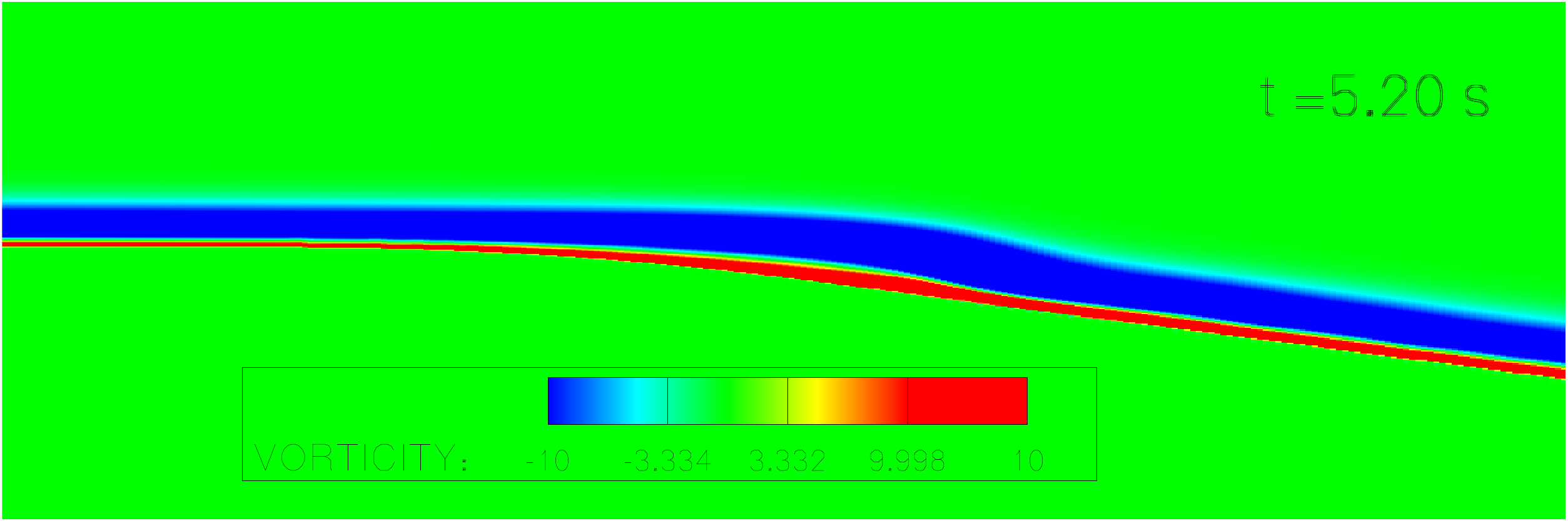}
		\end{subfigure}
	\vspace{2mm}
		\begin{subfigure}{0.22\textwidth}
			\centering
			\includegraphics[width=1\linewidth]{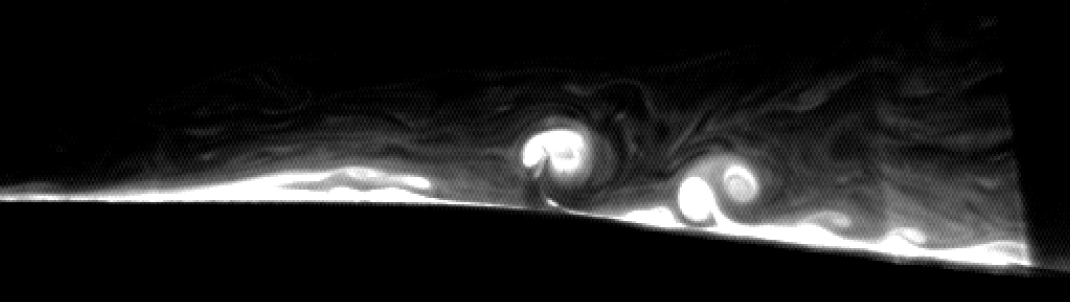}
		\end{subfigure}%
		\begin{subfigure}{0.22\textwidth}
			\centering
			\includegraphics[width=0.9\linewidth]{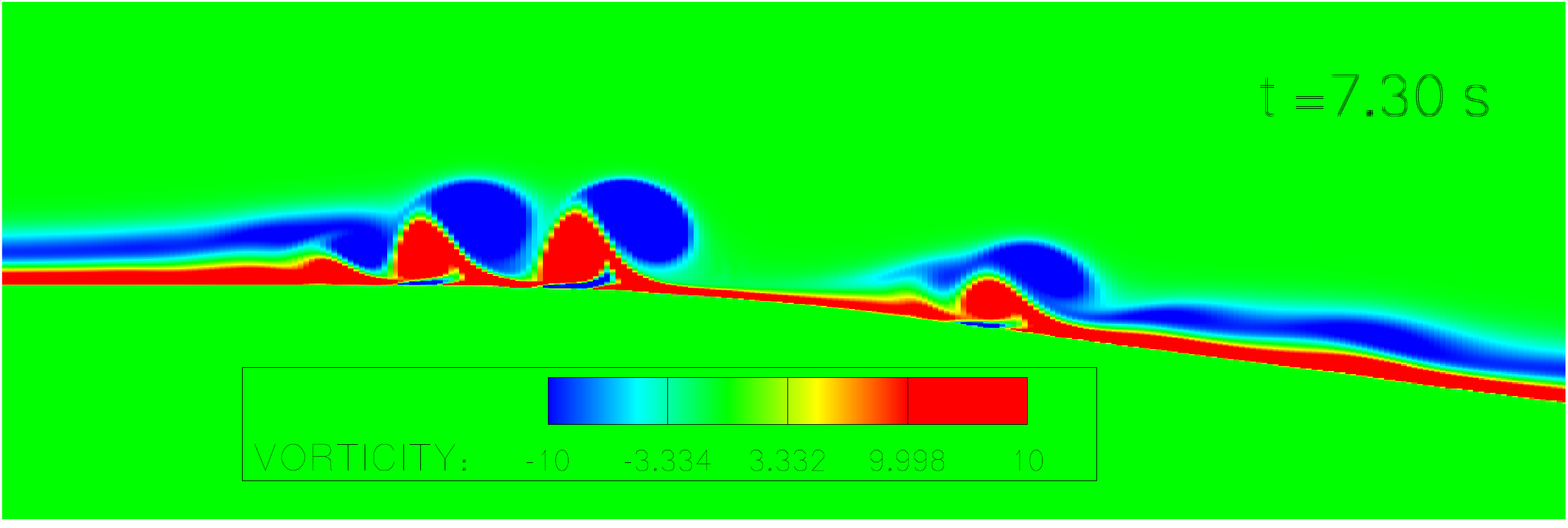}
		\end{subfigure}
		\begin{subfigure}{0.22\textwidth}
			\centering
			\includegraphics[width=\linewidth]{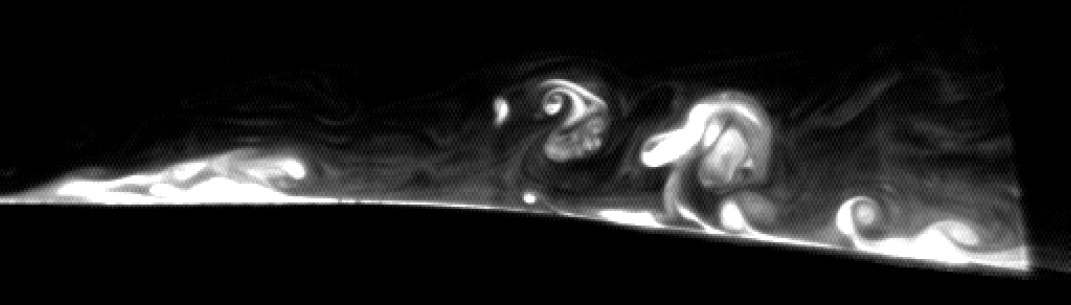}
		\end{subfigure}%
	\vspace{2mm}
		\begin{subfigure}{0.22\textwidth}
			\centering
			\includegraphics[width=0.9\linewidth]{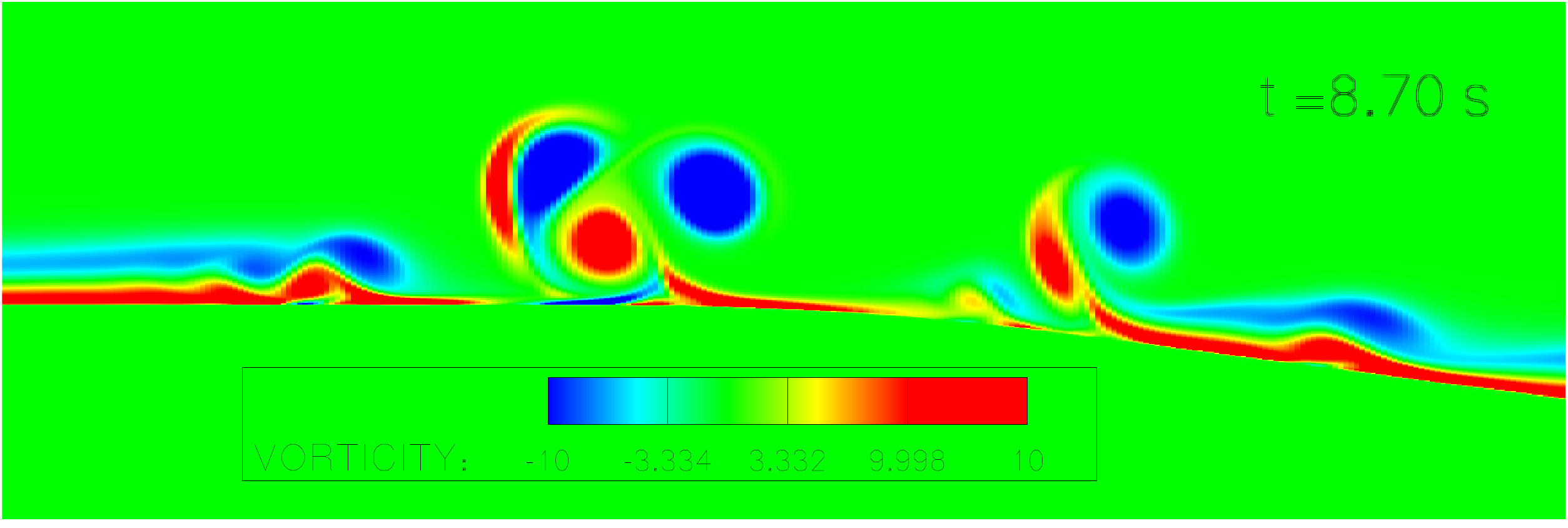}
		\end{subfigure}
				\vspace{2mm}
		\begin{subfigure}{0.22\textwidth}
			\centering
			\includegraphics[width=1\linewidth]{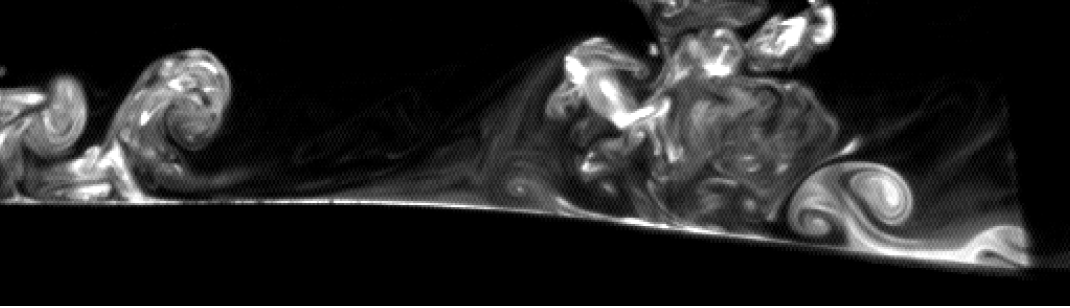}
		\end{subfigure}%
		\begin{subfigure}{0.22\textwidth}
			\centering
			\includegraphics[width=0.9\linewidth]{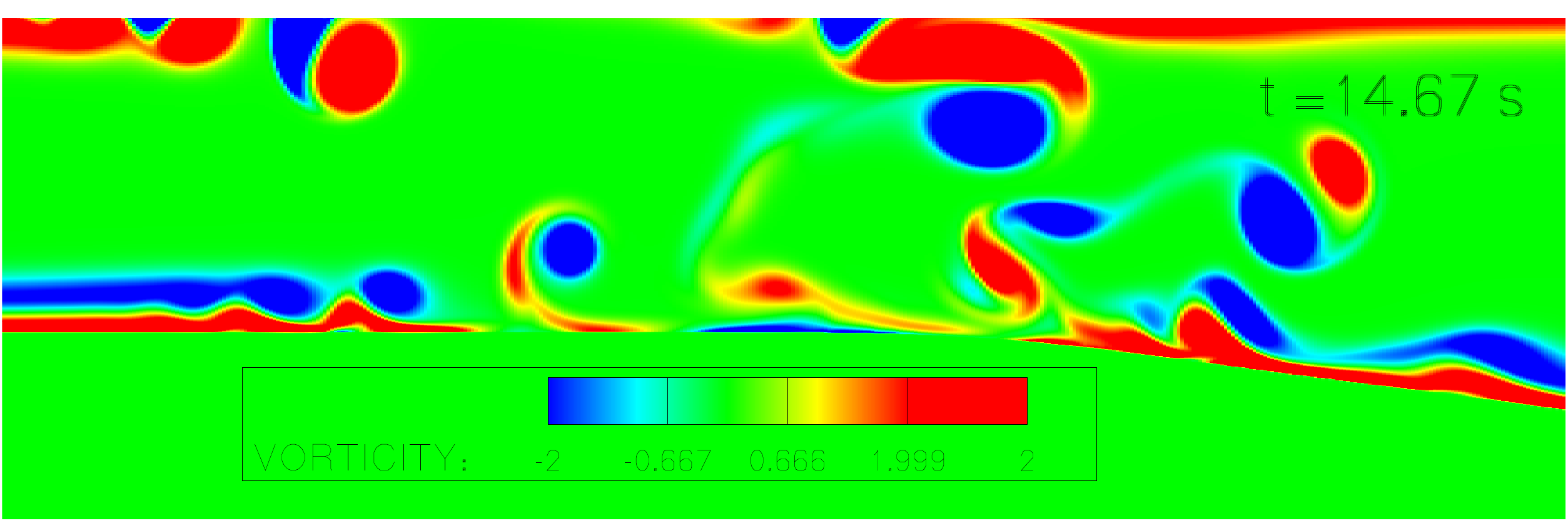}
		\end{subfigure}
		\caption[Case 2 Contour Plot] {CASE 2 $t_0=0.6 s, t_1=3.5 s, t_2=6.5 s, U_0=13.72 cm/s, Re_{\delta} \simeq 692 $ \\
			COMPARISON BETWEEN EXPERIMENTAL RESULT FROM DAS ET. AL \cite{das2016instabilities}  AND SIMULATION VORTICITY CONTOUR PLOT }
		\label{fig:case3}	
		\vspace{-10mm}
\end{figure}}

{\begin{figure}[h]
		\centering
		\begin{subfigure}{0.22\textwidth}
			\centering
			\includegraphics[width=1\linewidth]{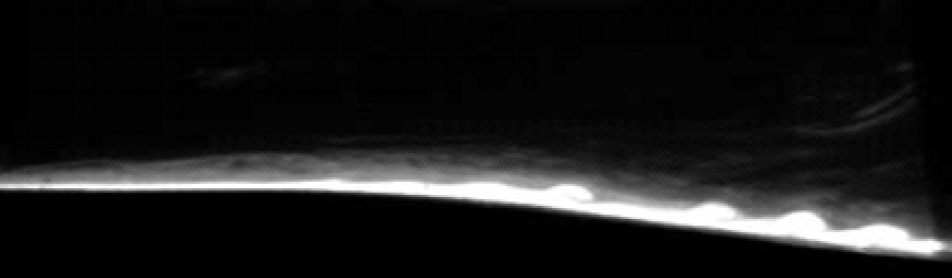}
		\end{subfigure}%
		\begin{subfigure}{0.25\textwidth}
			\centering
			\includegraphics[width=1\linewidth]{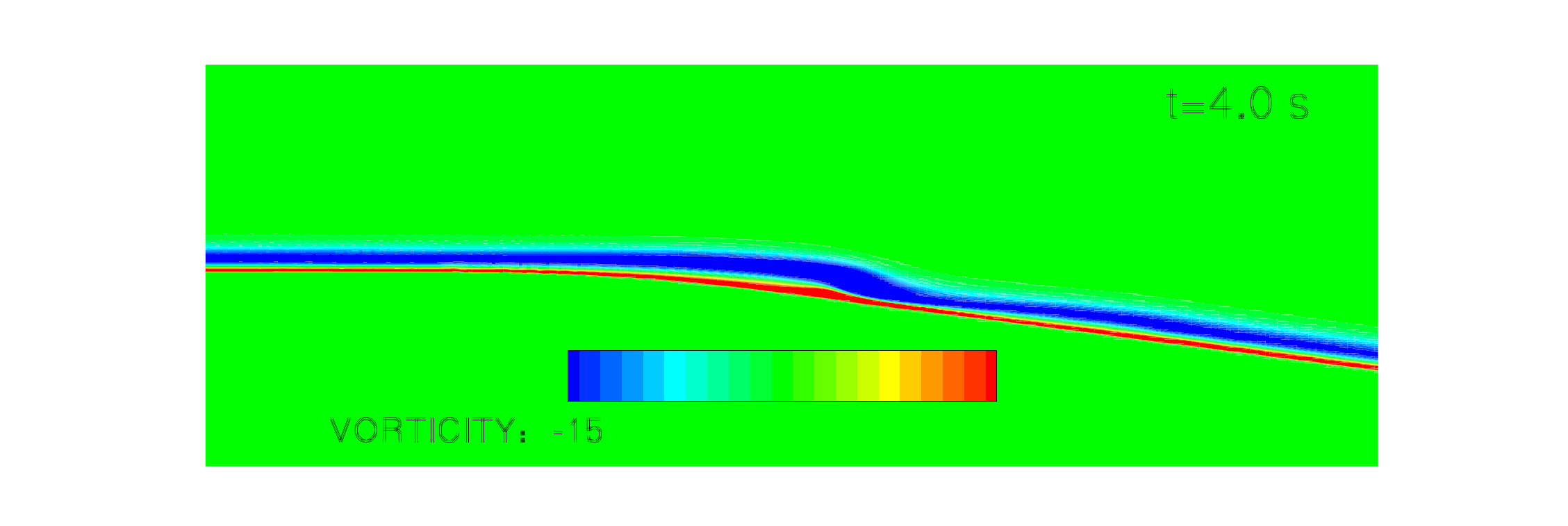}
		\end{subfigure}
		\begin{subfigure}{0.22\textwidth}
			\centering
			\includegraphics[width=1\linewidth]{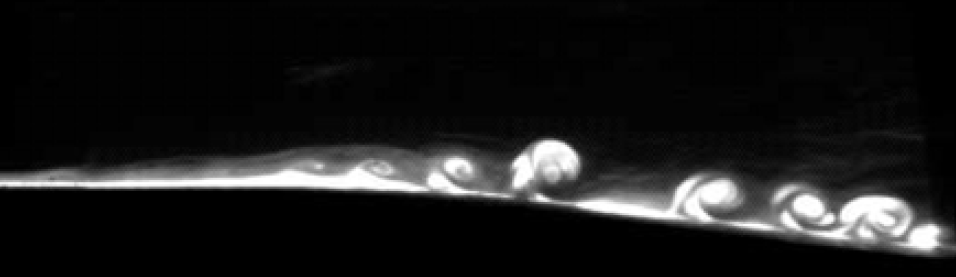}
		\end{subfigure}%
		\begin{subfigure}{0.25\textwidth}
			\centering
			\includegraphics[width=1\linewidth]{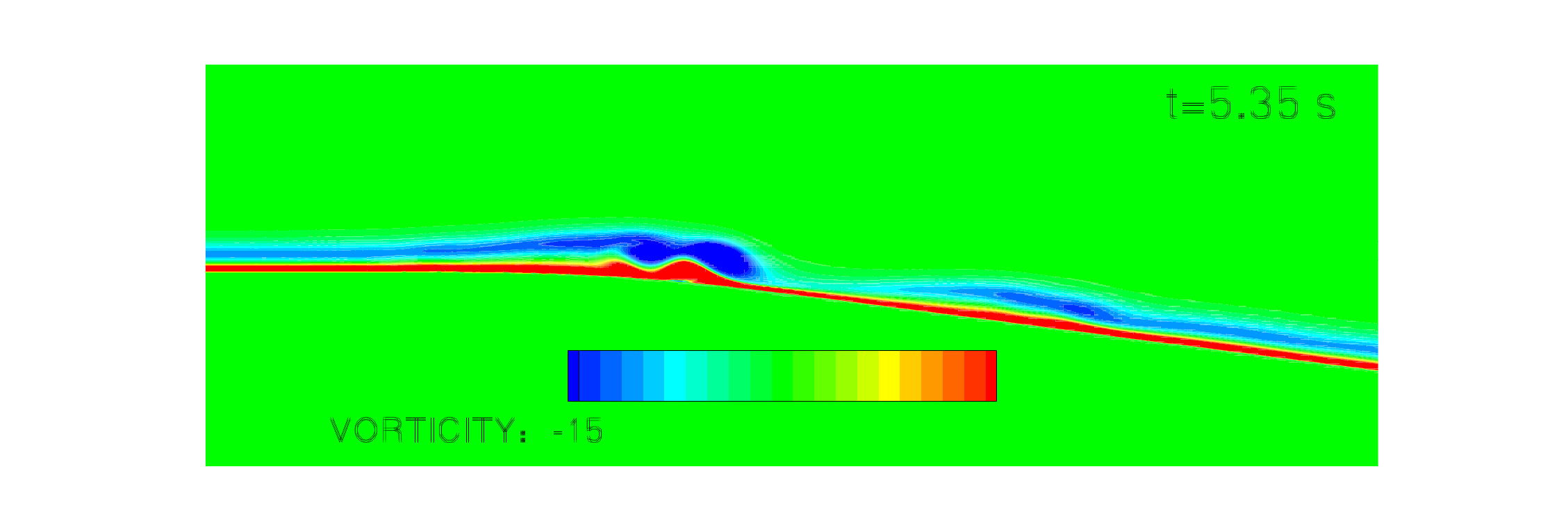}
		\end{subfigure}		
		\begin{subfigure}{0.22\textwidth}
			\centering
			\includegraphics[width=1\linewidth]{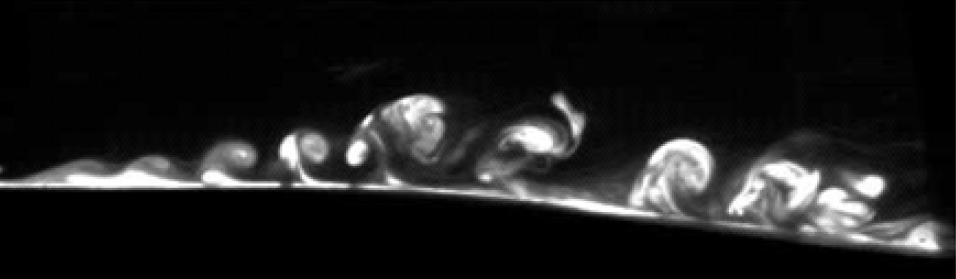}
		\end{subfigure}%
		\begin{subfigure}{0.25\textwidth}
			\centering
			\includegraphics[width=1\linewidth]{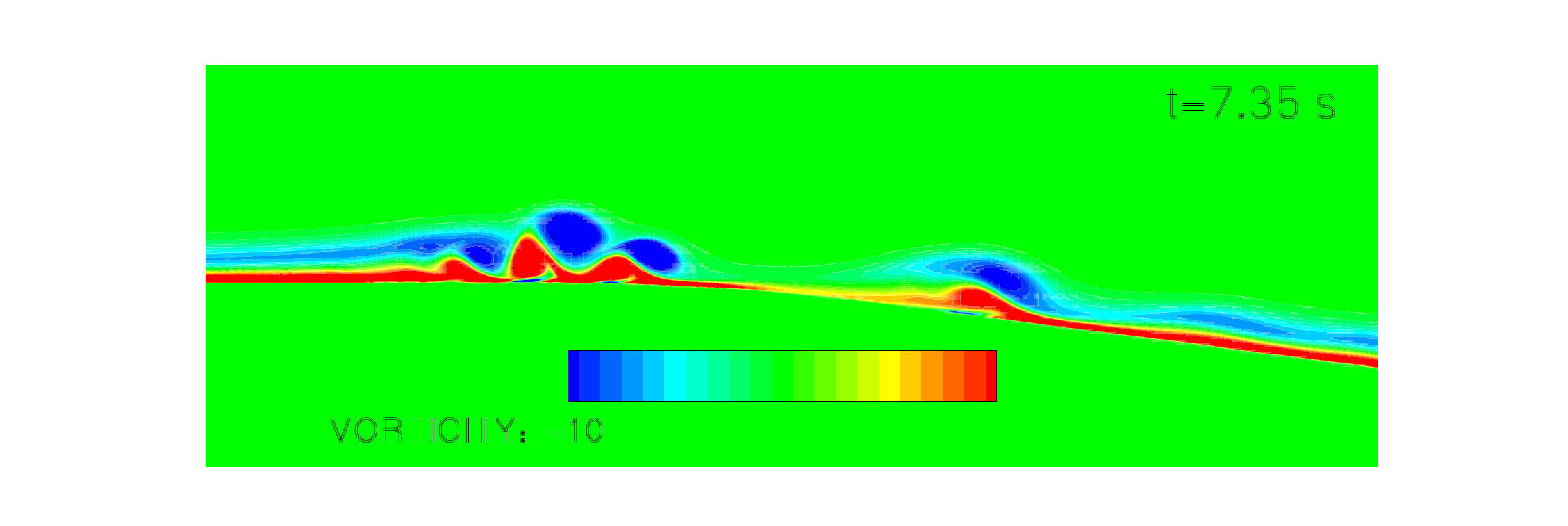}
		\end{subfigure}
		\begin{subfigure}{0.22\textwidth}
			\centering
			\includegraphics[width=1\linewidth]{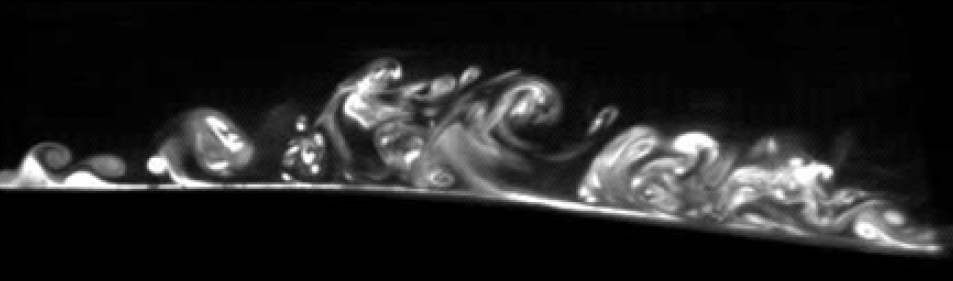}
		\end{subfigure}%
		\begin{subfigure}{0.25\textwidth}
			\centering
			\includegraphics[width=1\linewidth]{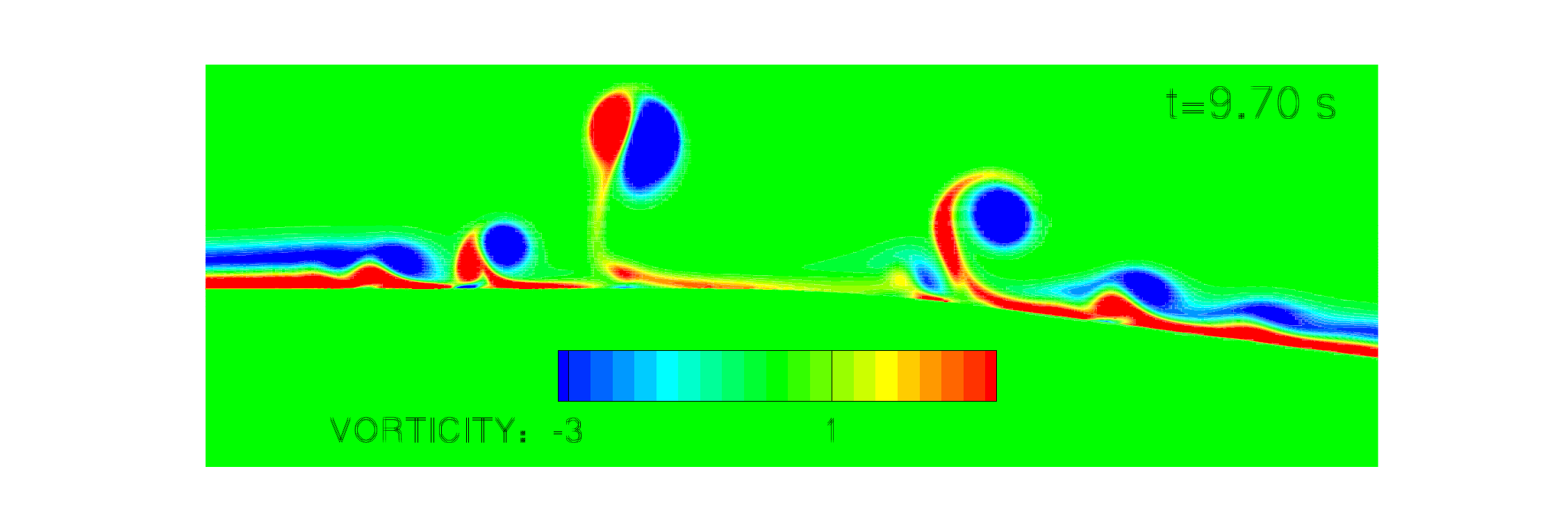}
		\end{subfigure}
		\caption[Case 3 Contour Plot] {CASE 3 $t_0=0.8 s, t_1=1.0 s, t_2=5.0 s, U_0=18.30 cm/s, Re_{\delta} \simeq 737 $ \\
	COMPARISON BETWEEN EXPERIMENTAL RESULT FROM DAS ET. AL. \cite{das2016instabilities}  AND SIMULATION VORTICITY CONTOUR PLOT }
		\label{fig:case7_1}
\end{figure}}
{\begin{figure}[h]
		\centering
		\begin{subfigure}{0.22\textwidth}
			\centering
			\includegraphics[width=1\linewidth]{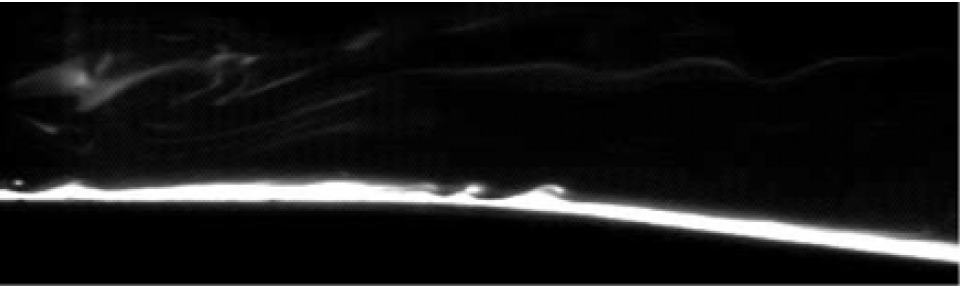}
		\end{subfigure}%
		\begin{subfigure}{0.25\textwidth}
			\centering
			\includegraphics[width=1\linewidth]{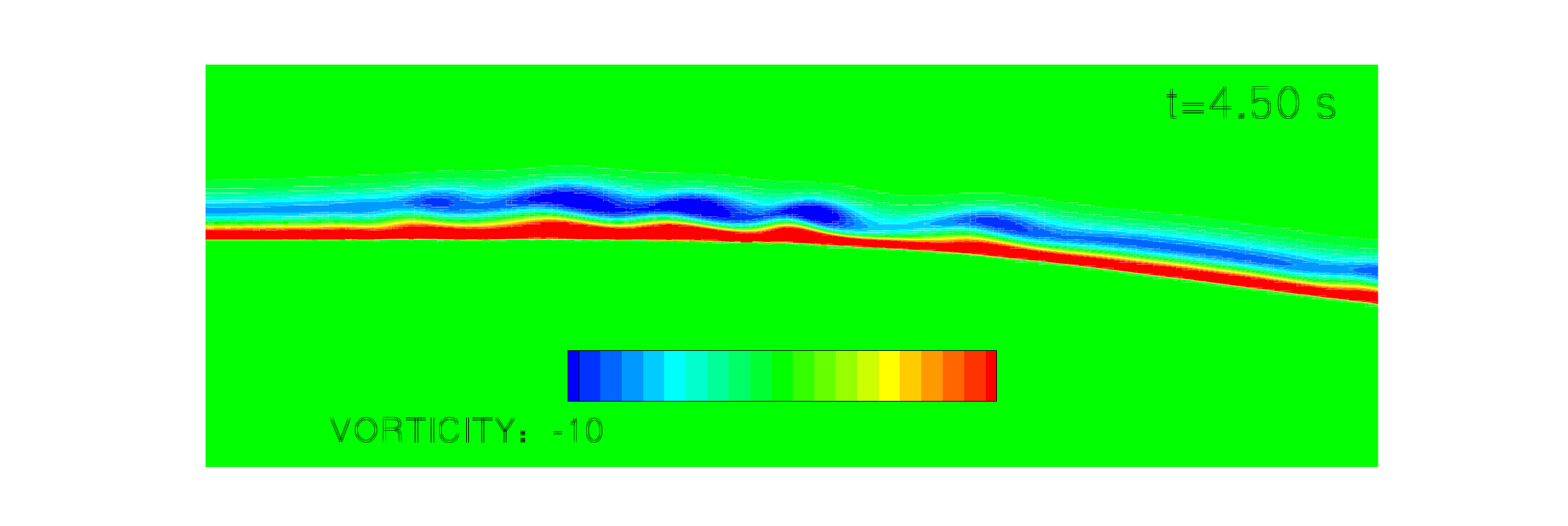}
		\end{subfigure}
		\begin{subfigure}{0.22\textwidth}
			\centering
			\includegraphics[width=1\linewidth]{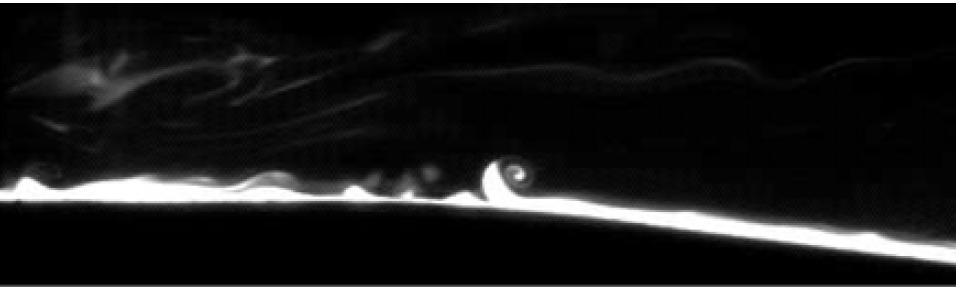}
		\end{subfigure}%
		\begin{subfigure}{0.25\textwidth}
			\centering
			\includegraphics[width=1\linewidth]{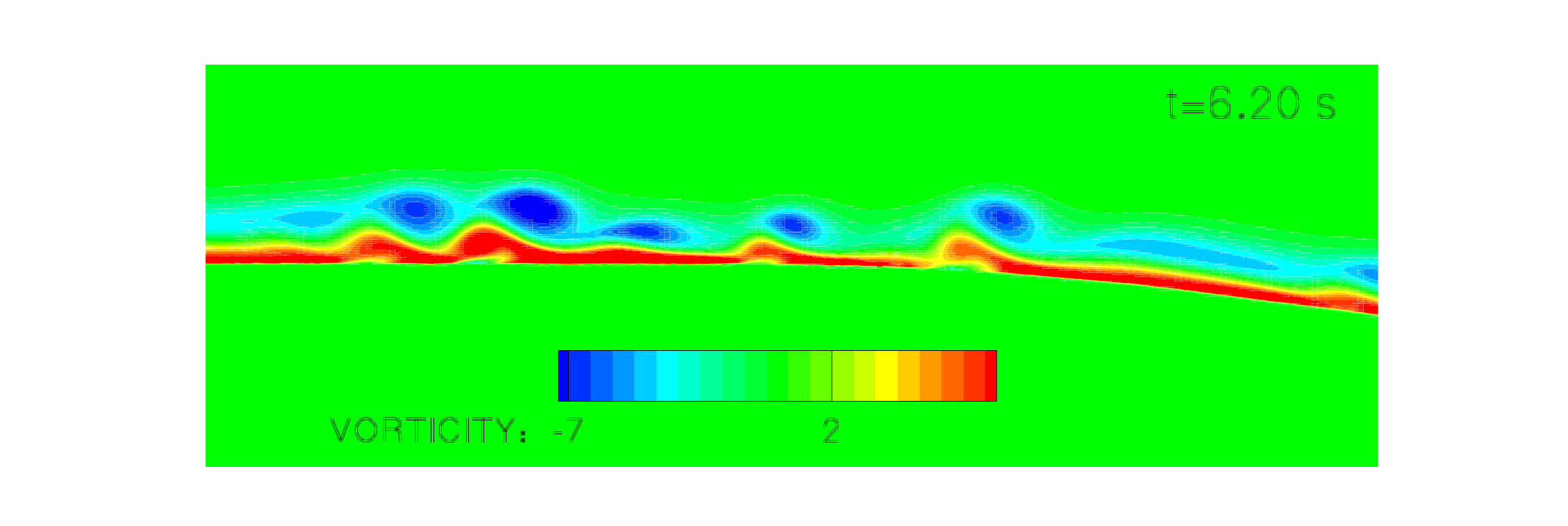}
		\end{subfigure}
		\begin{subfigure}{0.22\textwidth}
			\centering
			\includegraphics[width=1\linewidth]{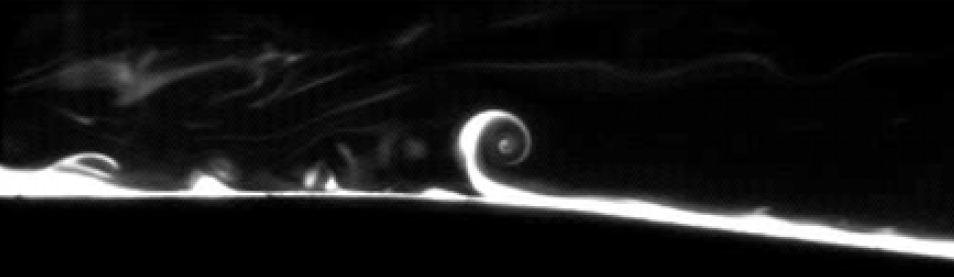}
		\end{subfigure}%
		\begin{subfigure}{0.25\textwidth}
			\centering
			\includegraphics[width=1\linewidth]{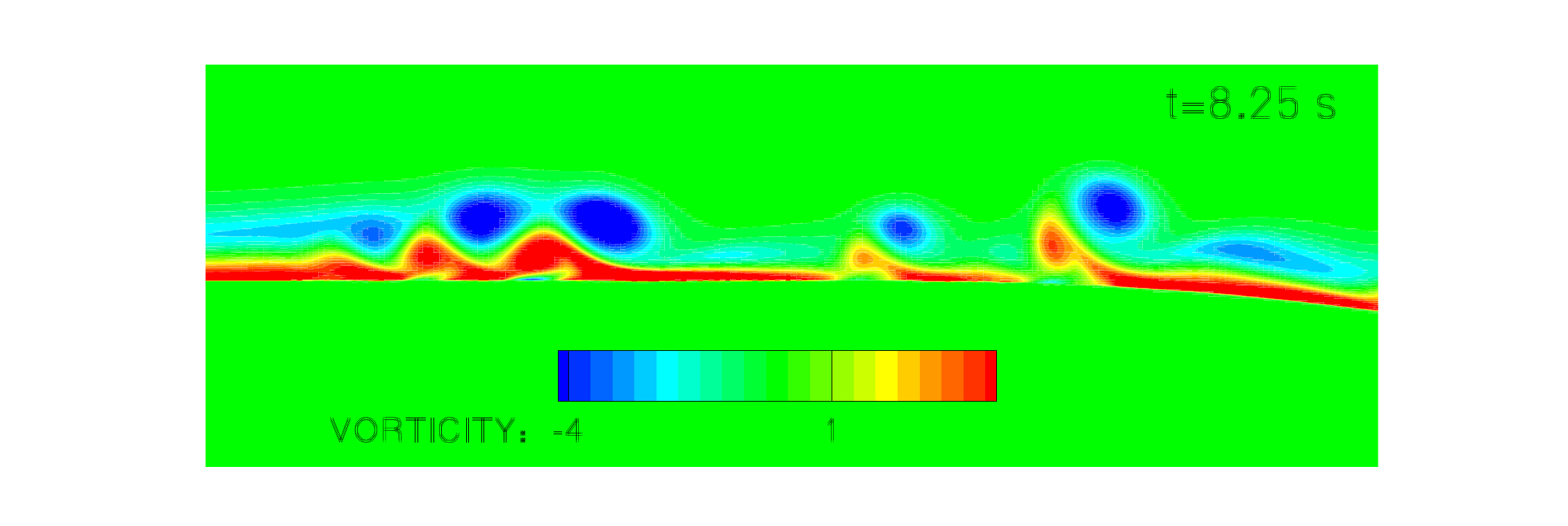}
		\end{subfigure}		
		\begin{subfigure}{0.22\textwidth}
			\centering
			\includegraphics[width=1\linewidth]{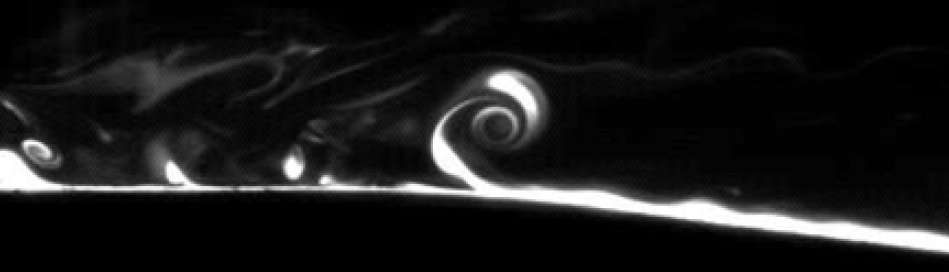}
		\end{subfigure}%
		\begin{subfigure}{0.25\textwidth}
			\centering
			\includegraphics[width=1\linewidth]{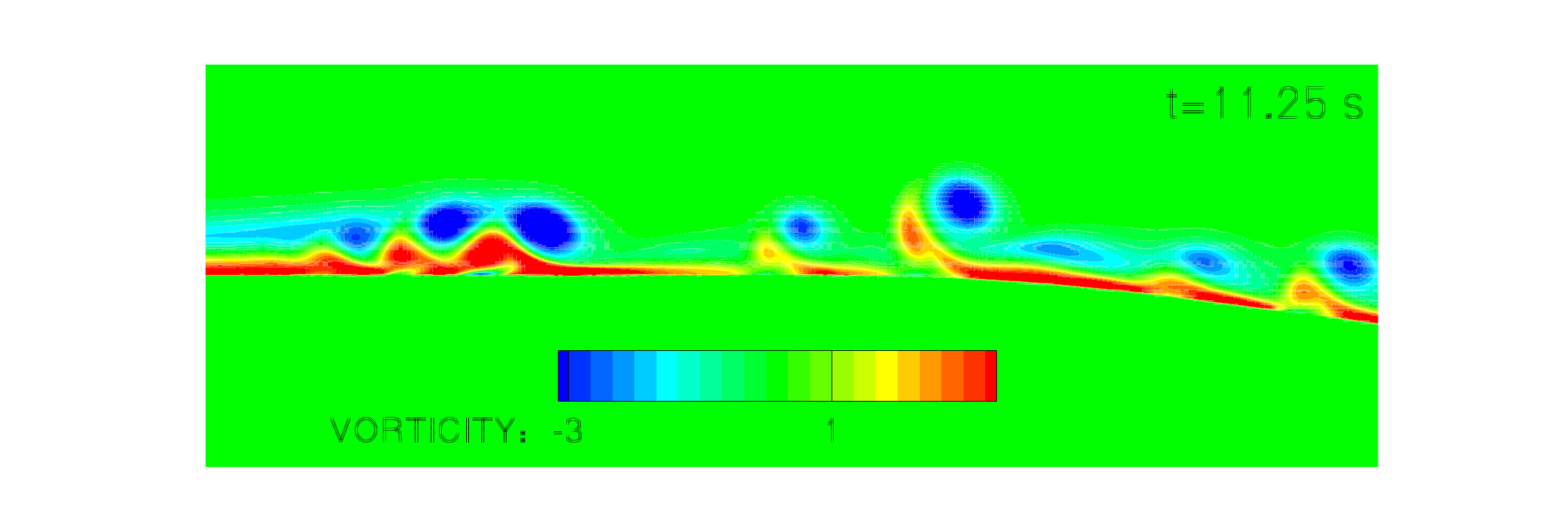}
		\end{subfigure}
			
		\caption[Case 4  Contour Plot] {CASE 4  $t_0=0.6 s, t_1=2.0 s, t_2=8.0 s, U_0=13.72 cm/s, Re_{\delta} \simeq 805 $ \\
		COMPARISON BETWEEN EXPERIMENTAL RESULT FROM DAS ET. AL \cite{das2016instabilities}  AND SIMULATION VORTICITY CONTOUR PLOT }
		\label{fig:case2_1}
		\vspace{-12mm}														
\end{figure}}

{\begin{figure}[h]
		\centering
		\begin{subfigure}{0.22\textwidth}
			\centering
			\includegraphics[width=1\linewidth]{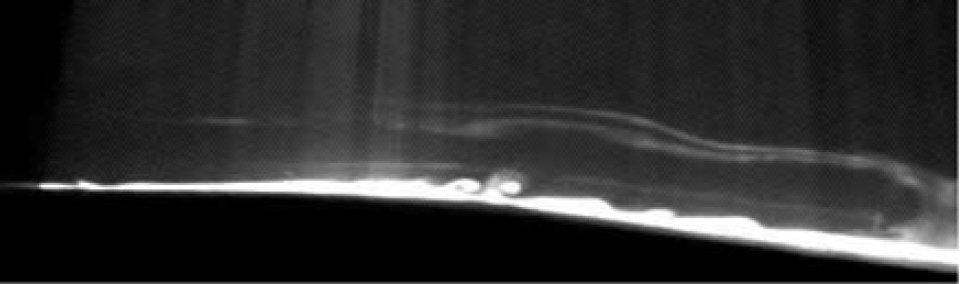}
		\end{subfigure}%
		\begin{subfigure}{0.25\textwidth}
			\centering
			\includegraphics[width=1\linewidth]{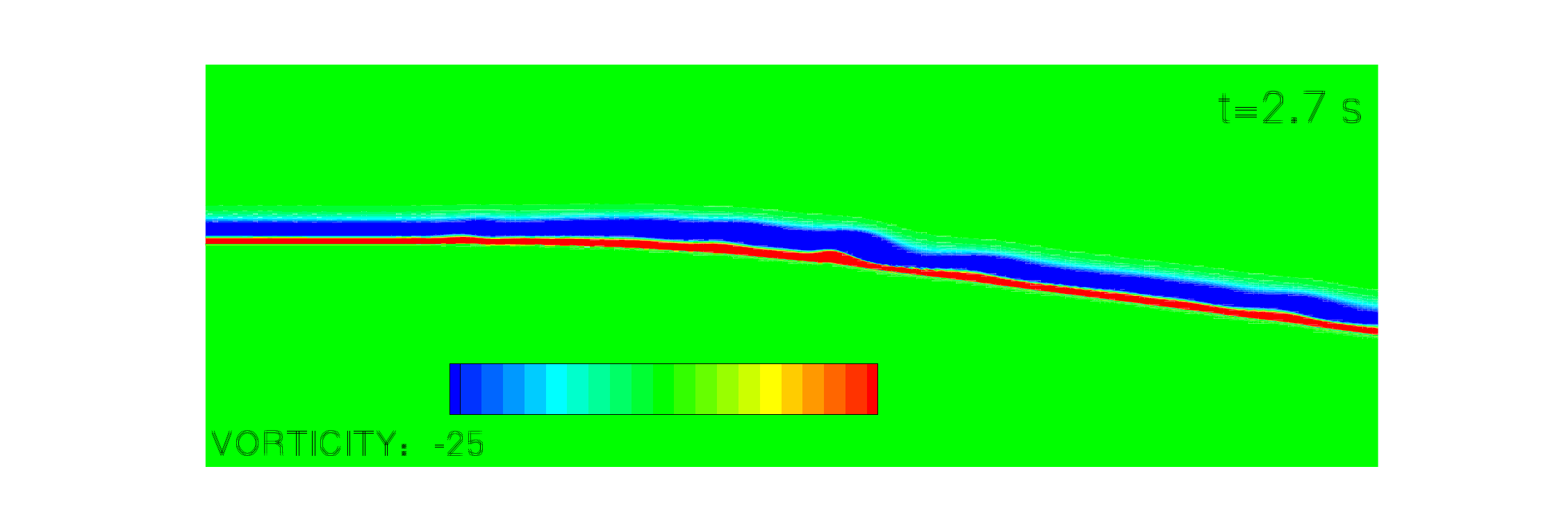}
		\end{subfigure}
		\begin{subfigure}{0.22\textwidth}
			\centering
			\includegraphics[width=1\linewidth]{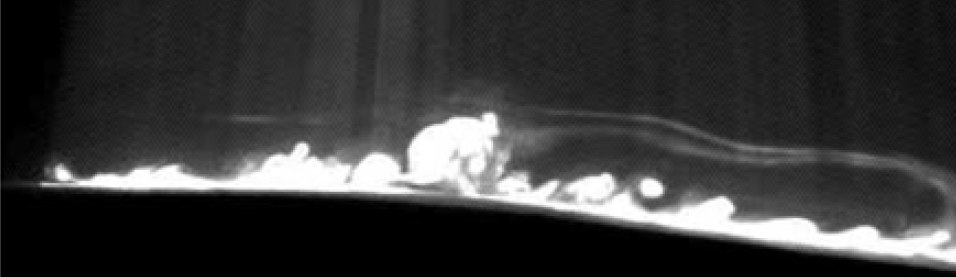}
		\end{subfigure}%
		\begin{subfigure}{0.25\textwidth}
			\centering
			\includegraphics[width=1\linewidth]{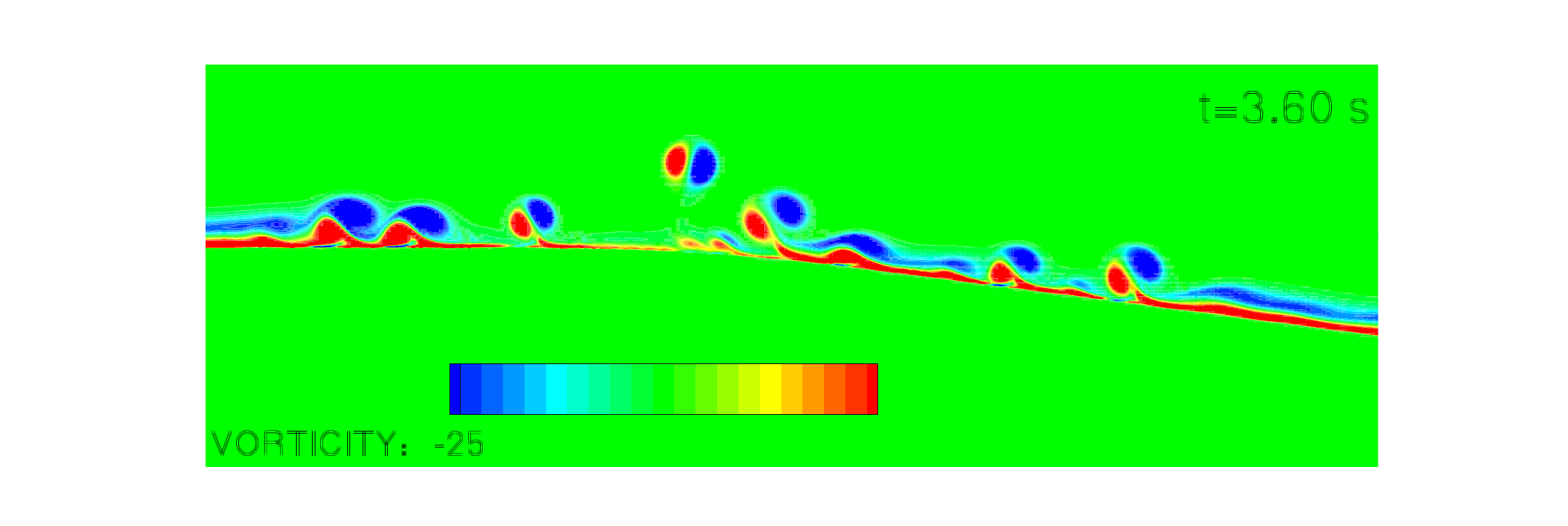}
		\end{subfigure}
		\begin{subfigure}{0.22\textwidth}
			\centering
			\includegraphics[width=1\linewidth]{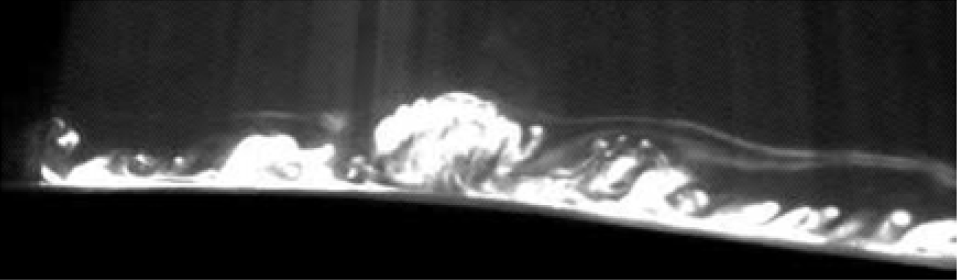}
		\end{subfigure}%
		\begin{subfigure}{0.25\textwidth}
			\centering
			\includegraphics[width=1\linewidth]{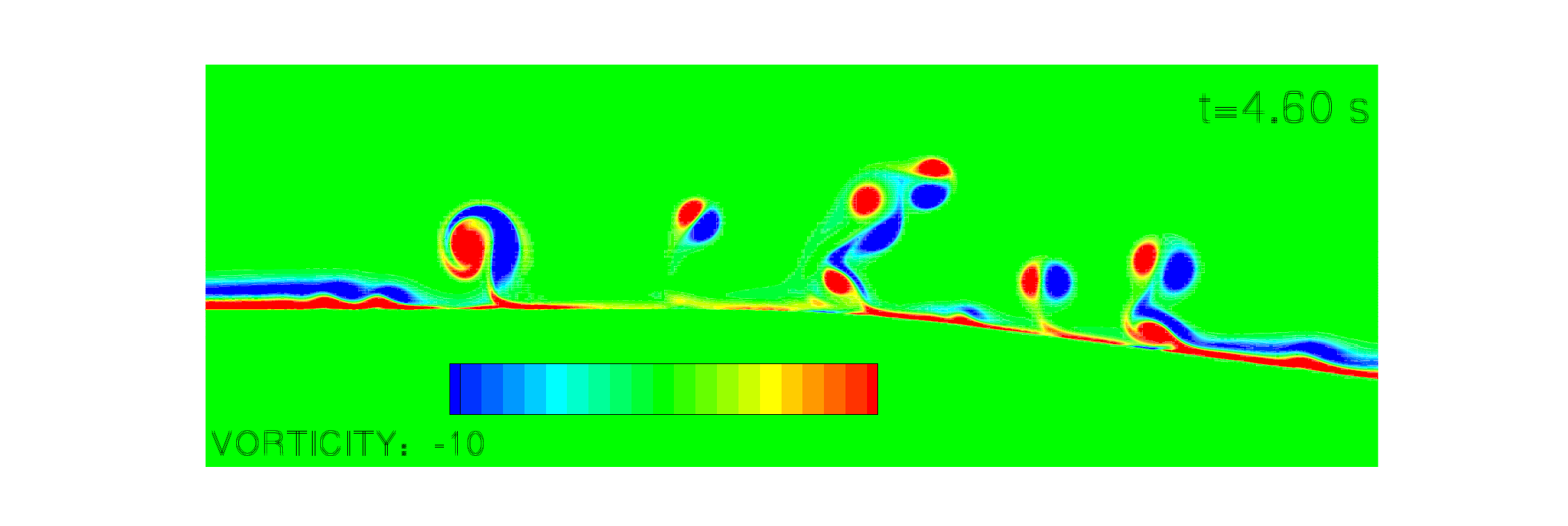}
		\end{subfigure}	
		\begin{subfigure}{0.22\textwidth}
			\centering
			\includegraphics[width=1\linewidth]{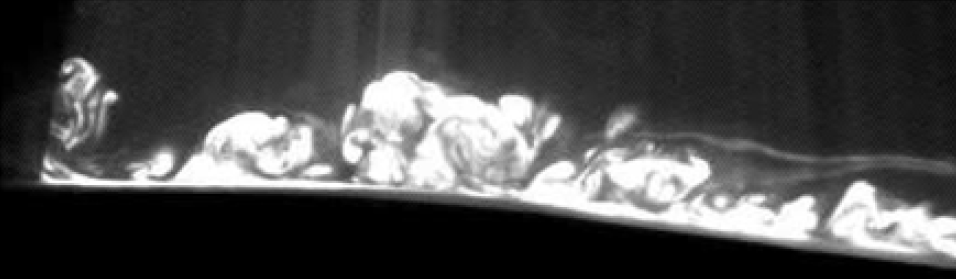}
		\end{subfigure}%
		\begin{subfigure}{0.25\textwidth}
			\centering
			\includegraphics[width=1\linewidth]{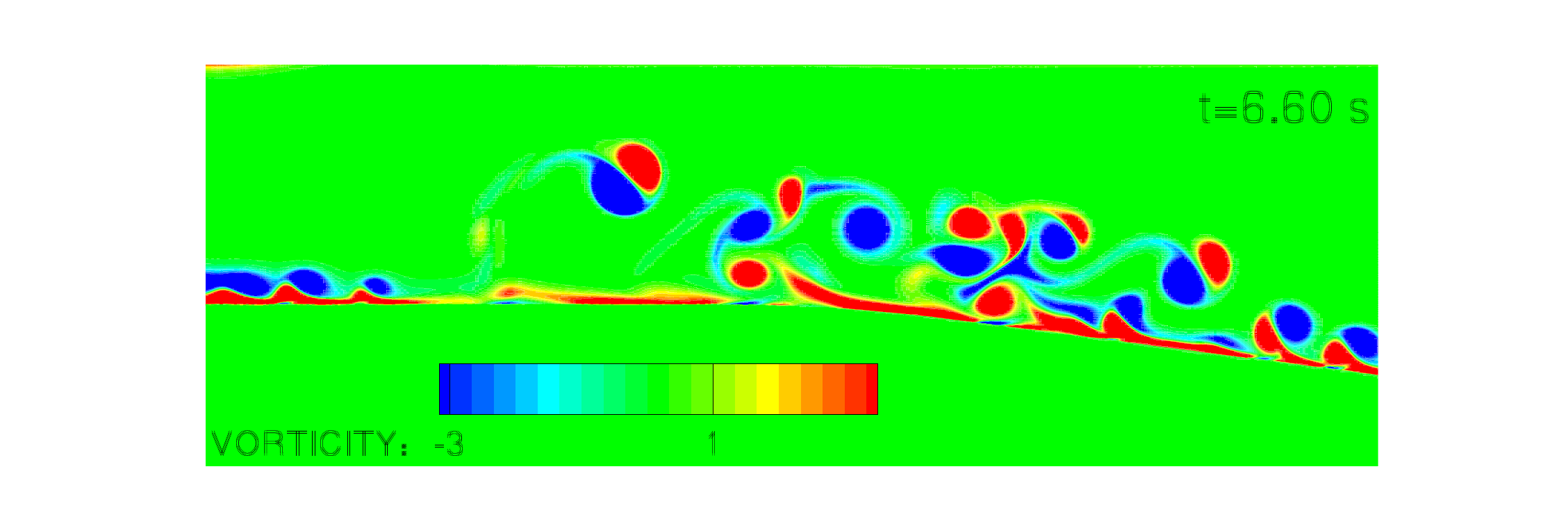}
		\end{subfigure}
		\caption[Case 5 Contour Plot (1)] {Case 5 $t_0=1.2 s, t_1=2.0 s, t_2=2.83 s, U_0=27.45 cm/s, Re_{\delta} \simeq 1355 $ \\
		COMPARISON BETWEEN EXPERIMENTAL RESULT FROM DAS ET. AL \cite{das2016instabilities}  AND SIMULATION VORTICITY CONTOUR PLOT }
		\label{fig:case15_1}
		\vspace{-13mm}												
\end{figure}}


In case 1, the deceleration is 2.29$m/s^2$ which is least among the all cases. Flow separation ocurred figure(\ref{fig:ts1}) in middle of the deceleration phase which is at 2.5 seconds. First vortex formed is at 4.8 seconds as shown in figure(\ref{fig:tv1}) whereas, first vortex seen by experiment was about 5.2 seconds. For this case, deceleration rate is least and in deceleration part because of adverse pressure gradient, flow near the wall is in reverse direction. Instabilities in free shear layer because of velocities in opposite direction could be the Kelvin-Helmholtz instability. Vortex which are formed at inlined section, are traveling in reverse direction.

In Case 2, acceleration of the flow is higher than the deceleration. Maximum velocity ($U_0$) attained 13.72 m/s at Reynolds Number is around 1095. Boundary layer thickness ($\delta_s$) is found to be 8.01mm. Separation of boundary layer occurred at end of constant velocity phase which is at 3.5 s as shown in Figure(\ref{fig:ts1}). First vortex was found during deceleration part itself. It was formed around 5.50 seconds as show in Figure(\ref{fig:tv2}). First vortex will move towards the flat section of the channel and number of vortex are also growing. In this case, vortex are traveled to flat section because of reverse flow are started to formed at the inclined section of the channel. First vortex is formed during deceleration part hence can be the Kelvin-Helmholtz instability.

The thing which makes case 3 special is the constant velocity piston motion given is merely for 0.2 seconds which is least among the all the cases. Maximum velocity ($U_0$) attained 18.30 m/s at Reynolds Number is around 737 and Reynold's number found by S.P Das \cite{das2016instabilities} is to be 719. Velocity profiles for x-direction at acceleration, constant velocity, deceleration and zero velocity are shown in figure(\ref{fig:profile1}). Boundary layer thickness at time of separation is 4.95 mm. Time of separation ($t_s$) is found to be 1.70 seconds. First vortex is formed during the deceleration period which is $t_v=$3.50 seconds as shown in figure(\ref{fig:tv1}). First vortex was observed in inclined section of the channel and it moved in reverse direction of flow due to adverse pressure gradient.

In Case 4 , acceleration and deceleration rates are moderate as compared with the other cases. Maximum velocity ($U_0$) attained 13.72 m/s at Reynolds Number is around 713. This is lowest Reynold's number case for which simulation is done. Boundary layer thickness ($\delta_s$) is found to be 5.41mm.  Separation of the boundary layer is seen at the beginning of deceleration phase. As shown in the figure(\ref{fig:ts2}) time of separation $t_s$ for Case 4 is 2.05 seconds. First vortex was seen well after the piston has stopped.  Most of the vortex formed in this case are on the flat section. It was observed that vortex formed at inclined sections are are moving in reverse direction of flow towards the flat section. Reverse flow is observed in both flat and diverging sections. From figure(\ref{fig:tv2}), it can be seen that first vortex is observed first vortex formed is around 4.3 seconds. 

Case 5 has the highest Reynold's number ($Re_{\delta s}=1434$), highest maximum velocity ($U_0$) and highest deceleration rate (d) among all other cases done. Separation has occurred at end of constant piston velocity at 2 seconds (figure(\ref{fig:ts2})) and first vortex is formed during deceleration motion of the piston (figure(\ref{fig:tv2})). It can be also seen that more than one vortex are formed within 2 seconds from figure(\ref{fig:tv2}). Velocity profiles in x-direction for all cases at 4 seconds is shown in figure(\ref{fig:profile2})

{\begin{figure}[htb!]
		\centering												
		\includegraphics[width=1\linewidth]{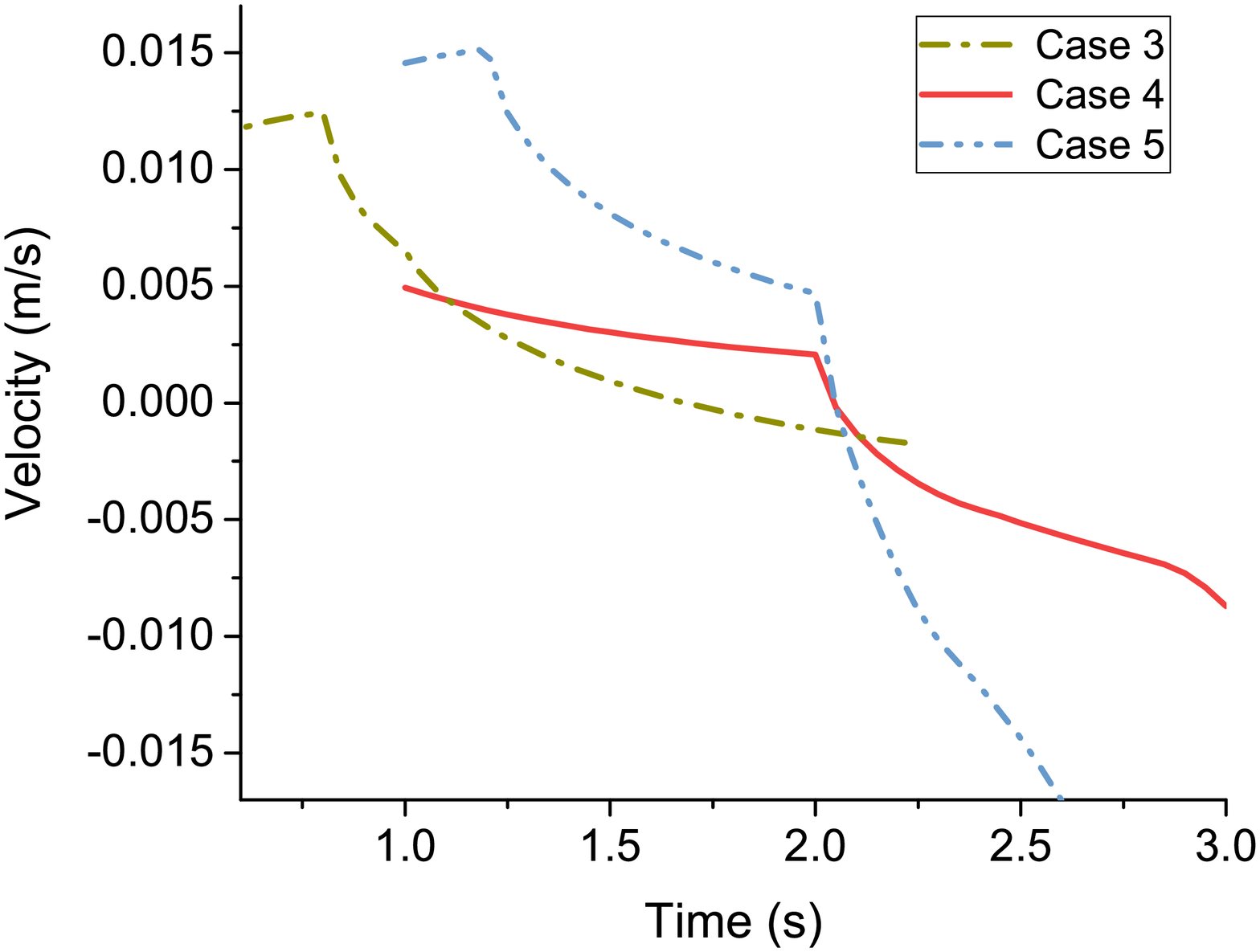}			
		\caption{VELOCITY IN X-DIRECTION VS TIME AT FIRST POINT OF SEPARATION}
		\label{fig:ts2}
		\vspace{-5mm}												
\end{figure}
	
{\begin{figure}[htb!]
			\centering												
			\includegraphics[width=1\linewidth]{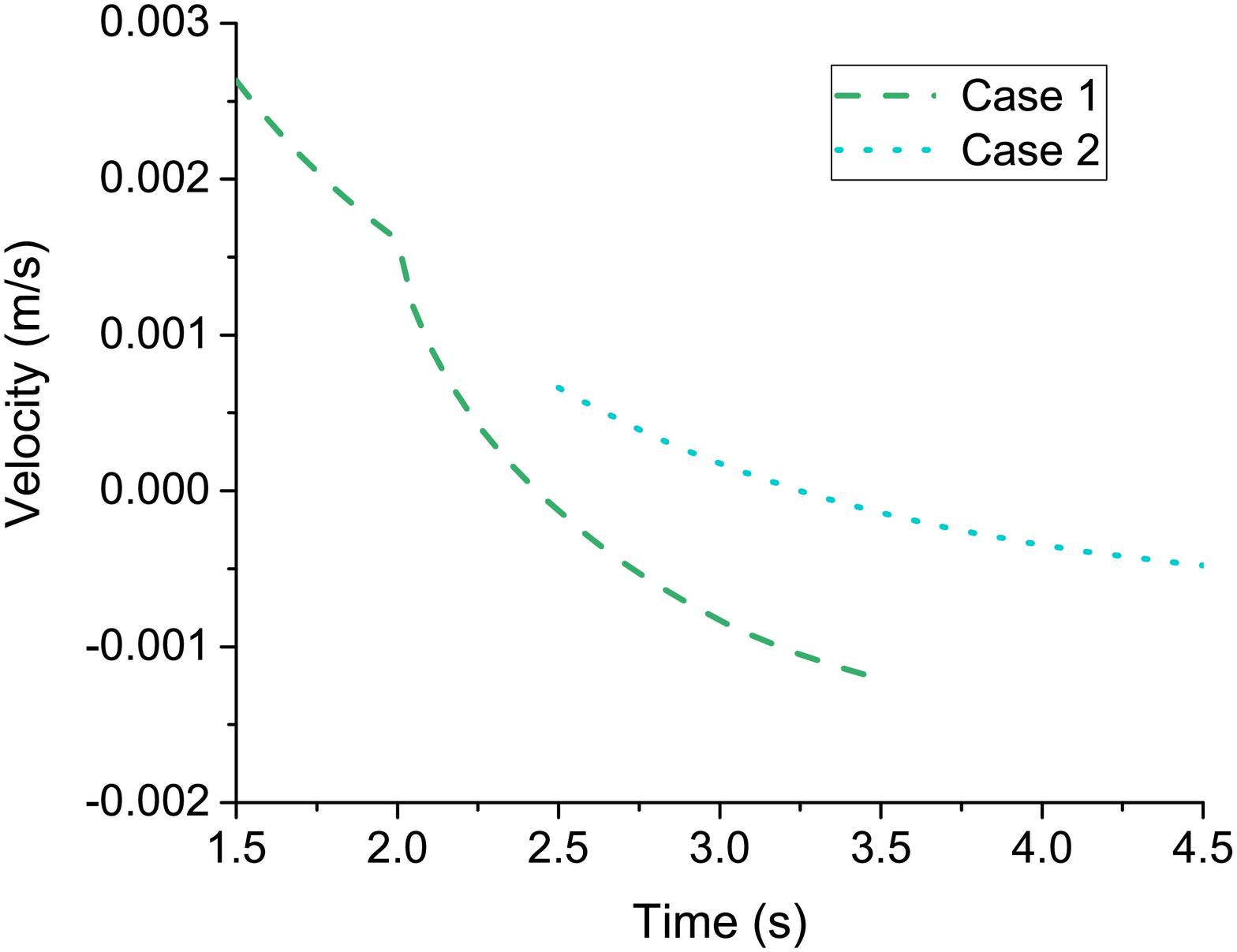}			
			\caption{VELOCITY IN X-DIRECTION VS TIME AT FIRST POINT OF SEPARATION}
			\label{fig:ts1}
			\vspace{-5mm}												
\end{figure}

{\begin{figure}[htb!]
		\centering												
		\includegraphics[width=1\linewidth]{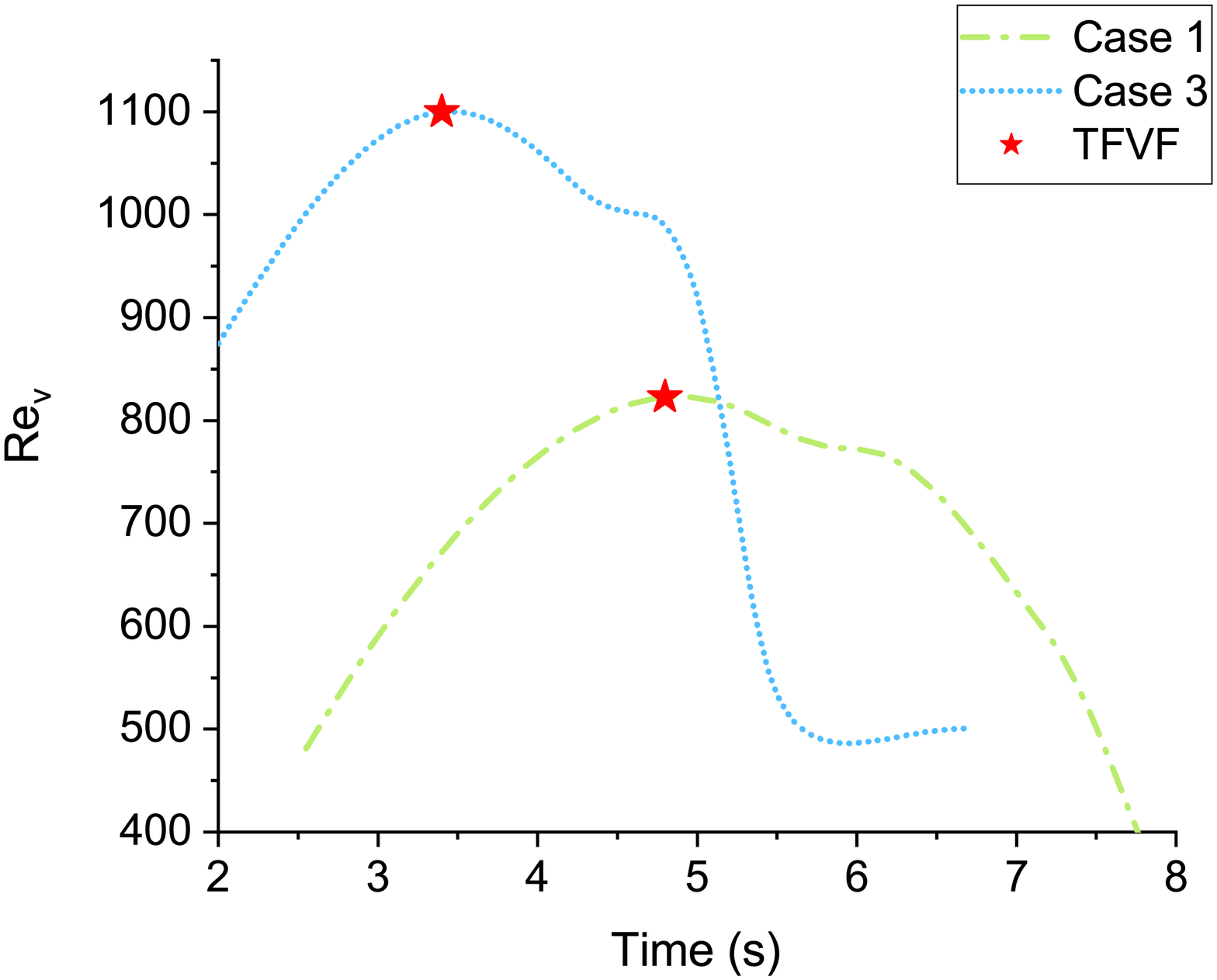}			
		\caption{TIME OF FIRST VORTEX FORMATION (TFVF)}
		\label{fig:tv1}
		\vspace{-5mm}												
	\end{figure}
	
	{\begin{figure}[htb!]
			\centering												
			\includegraphics[width=1\linewidth]{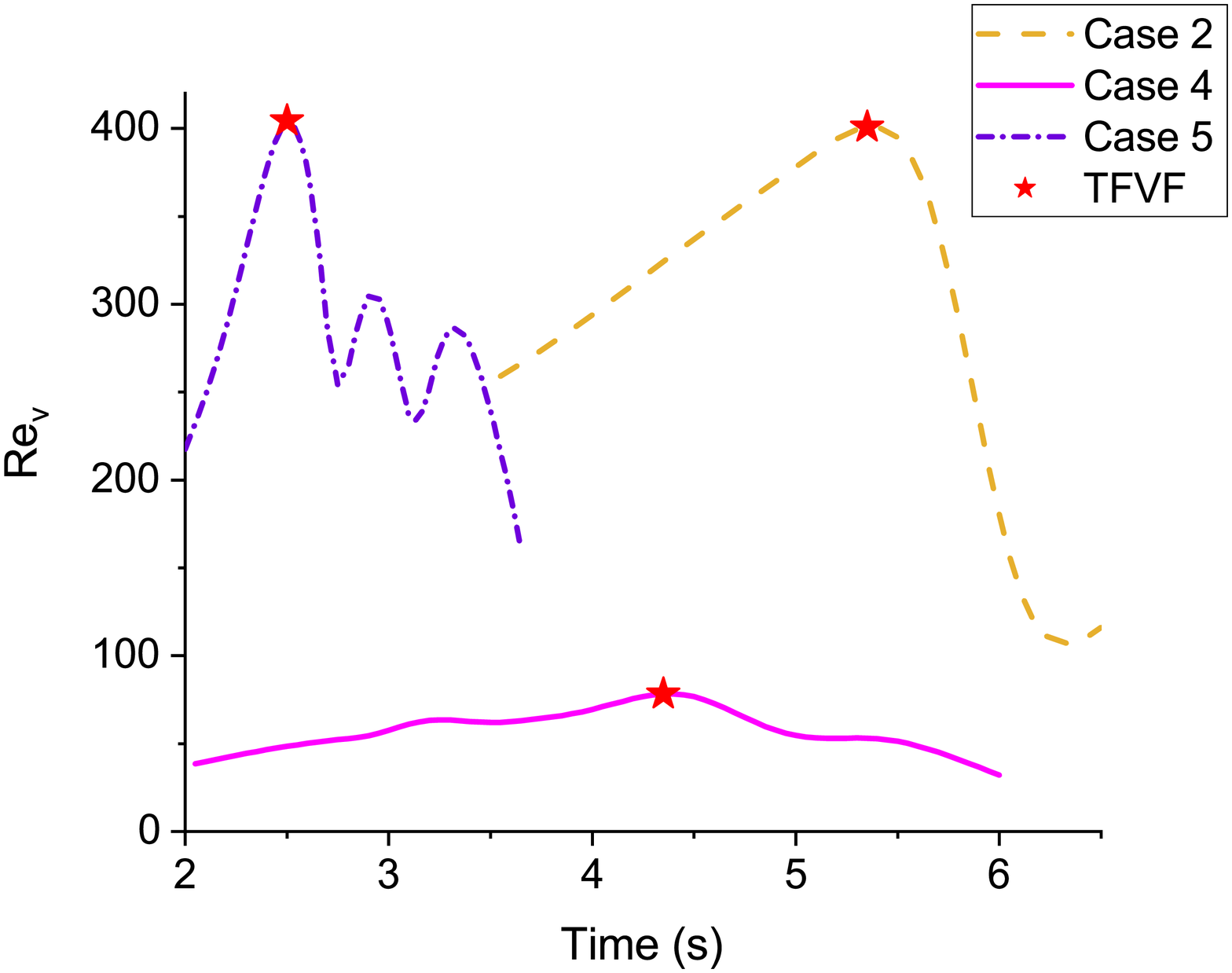}			
			\caption{TIME OF FIRST VORTEX FORMATION (TFVF)}
			\label{fig:tv2}
			\vspace{-5mm}												
		\end{figure}
	
{\begin{figure}[htb!]
		\centering									
		\includegraphics[width=1\linewidth]{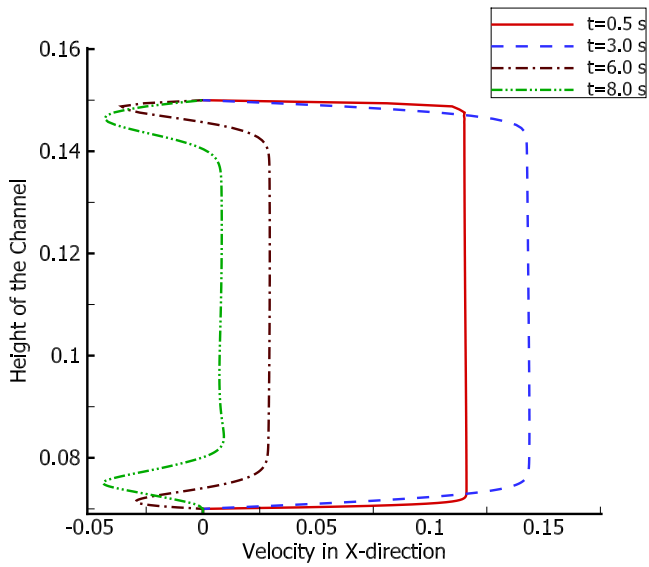}			
		\caption{VELOCITY PROFILES AT DIFFERENT TIMES FOR CASE 3}
		\label{fig:profile1}
		\vspace{-5mm}											
	\end{figure}	
{\begin{figure}[htb!]
		\centering									
		\includegraphics[width=1\linewidth]{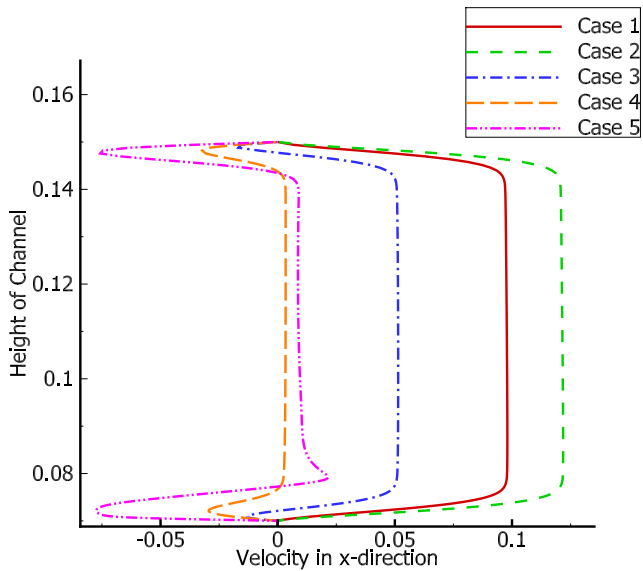}			
		\caption{VELOCITY PROFILES FOR ALL CASES AT T= 4 SECONDS}
		\label{fig:profile2}
		\vspace{-5mm}											
	\end{figure}
Vortex formation times are taken from time of separation to time of first vortex formation. It can be written as, Vortex formation time$ = t_v-t_s$. Time of vortex formation varies from 0.5 s to 3.05 s. Vortex formation times are normalized with velocity ($\Delta U_s$) and boundary layer thickness ($\delta_s$). Non-dimensional vortex formation time is plotted with Reynolds number ($Re_{\delta_s}$) as shown in figure(\ref{fig:nondim}).

	{\begin{figure}[htb!]
		\centering												
		\includegraphics[width=1\linewidth]{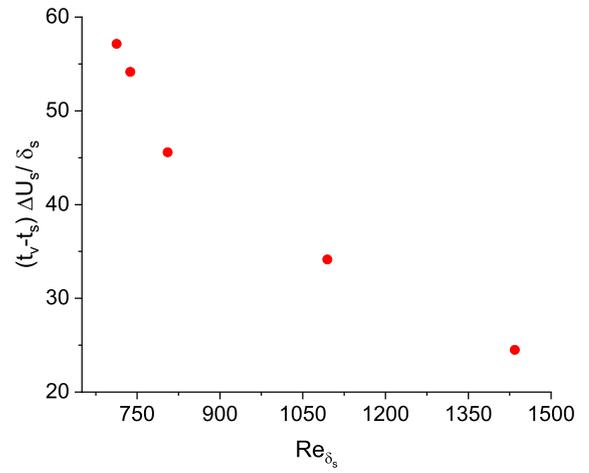}			
		\caption{ NON-MULTIDIMENSIONAL VORTEX FORMATION TIME WITH REYNOLDS NUMBER}
		\label{fig:nondim}
		\vspace{-5mm}												
	\end{figure}

\section*{CONCLUSION}
Unsteady boundary layer and associated instability charac- teristics are analysed using direct numerical simulations for five different transient inflow conditions. Qualitative and quantitative comparison with the experiments are performed by comparing spanwise vorticity, flow separation and vortex formation time. A relatively good agreement with the experiments are observed for all the cases. The formed vortices exhibits a wide range of dy- namics and the common feature observed for all the cases are the cat-eye like patterns similar to the Kelvin Helmholtz instability formed during the decelerating phase of the flow.

\bibliographystyle{ihmtc}
\bibliography{IHMTC-2019_template}

\end{document}